%% file: main.tex
\pdfoutput=1
\documentclass[amsfonts, amssymb, amsmath, nobibnotes,twocolumn,superscriptaddress,eqsecnum,floatfix]{revtex4-1}
\usepackage[english]{babel}
\usepackage[utf8]{inputenc}
\input{preamble}
\usepackage{wrapfig, subfig}
\usepackage{float}
\restylefloat{figure}
\usepackage[labelfont=bf]{caption}
\begin{document}
\title{Deterministic and Stochastic Research of \\ Cubic Population Model with Harvesting}

\author{Özgür Gültekin}
	\email[Corresponding author: ]{gultekino@yahoo.com}
\affiliation{Mimar Sinan Fine Arts University, Department of Statistics, Istanbul 34427, Turkey}
\affiliation{Mimar Sinan Fine Arts University, Department of Mathematics, Istanbul 34427, Turkey}

\author{Çağatay Eskin}
	\email[]{cagatay.eskin@metu.edu.tr}
\affiliation{Middle East Technical University, Department of Physics, Ankara 06800, Turkey}
    
\author{Esra Yazıcıoğlu}
	\email[]{yaziciogluesra@gmail.com}
\affiliation{Istanbul University, Department of Biology, Istanbul 34452, Turkey}


\begin{abstract}
A detailed examination of the effect of harvesting on a population has been carried out by extending the standard cubic deterministic model by considering a population under Allee effect with a quadratic function representing harvesting. Weak and strong Allee effect transitions, carrying capacity, and Allee threshold change according to harvesting is first discussed in the deterministic model. A Fokker Planck equation has been obtained starting from a Langevin equation subject to correlated Gaussian white noise with zero mean, and an Approximate Fokker Planck Equation has been obtained from a Langevin equation subject to correlated Gaussian colored noise with zero mean. This allowed to calculate the stationary probability distributions of populations, and thus to discuss the effects of linear and nonlinear (Holling type-II) harvesting for populations under Allee effect and subject to white and colored noises respectively.
\end{abstract}

\maketitle

\section{Introduction}
Population dynamics have a wide range of applications beyond the limits of ecology and biology \cite{murray,kot,Metapopulation,Mathematicalmodels}. For example, theoretical and practical tools for population dynamics are used even in areas that are not directly related to ecology such as astrophysics and plasma physics \cite{Diamond,Morel2014,Ross}. An important phenomenon in terms of the consequences of intraspecific cooperation in a population is the Allee effect. Allee effect was first described by Warder Clyde Allee in the 1930's \cite{Allee,Allee1938}. Despite its long history, it is still a current research topic in the fields of population dynamics, evolutionary biology and genetics \cite{Qin2017,Sun2016,Schreiber2003a,Cheptou2004}. Allee effect is defined as positive correlation between population density and mean individual fitness \cite{courchamp,Courchamp_inverse_density,What_is_allee_effect_stephens}. In the classical description of evolution of population dynamics, it is expected per capita growth rate to increase with the increasing population density. On the other hand, while the per capita growth rate increases with the increasing population density at large population densities, the decrease in the per capita growth rate with decreasing population density at low population densities means that the population shows Allee effect. If population always goes to extinction for average population size values below a threshold value, this is strong Allee effect. In the absence of such a critical average population size, if the Allee effect is observed although the per capita growth rate of population remains always positive, this is defined as weak Allee effect \cite{courchamp}. There are some mechanisms known that can cause Allee effect such as finding mate in low population densities being more difficult than in normal conditions \cite{Berec2001,Groom1998,Levitan2012}, prey groups with small number of individuals resulting in a weakened defense against predators \cite{Wittmer2005,Gascoigne2004}, population of some of the sea creatures decreasing while hunting does not decrease \cite{Frank2000} and limitation of evolutionary possibilities through the accumulation of harmful mutations in populations with low density \cite{Fischer2008}. As the evidence that low density population dynamics are determined under Allee effect \cite{Courchamp1999,Kuussaari1998,Angulo2007,Kramer2009} increased, interest and theoretical studies about the Allee effect have also increased in recent years. Populations showing Allee effect under stochastic effects are likely to extinct. Therefore, research on Allee effect supports the development of new strategies for population conservation \cite{akcakaya,Stephens1999,Ackleh2007}. Understanding the effects of foreign species invaded a new region as a result of human activity is also an important problem in terms of conservation biology \cite{Ackleh2007,Sasmal2017}. \par

Although there are studies considering the harvesting effect in deterministic population models for a single species \cite{Constant_harvest_brauer,Logistic_type_with_harvesting_liu}, the harvesting effect has not been studied sufficiently in stochastic models. When stochastic population models are expanded to include the harvesting effect, the systems described by the model also vary \cite{kot}. In addition, the function of the harvesting to reduce population density may result in populations showing weak Allee effects start to show strong Allee effect. It can be expected that this may cause population to extinct faster than the population without harvesting. Harvesting a population showing Allee effect provides the opportunity to address problems such as the management of renewable resources as an optimal harvesting problem. Deterministic analysis of such an optimum harvesting problem has been made by Srinivasu and Kumar \cite{Srinivasu2014}. On the other hand, there are no comprehensive stochastic studies in the literature in which a cubic population model showing Allee effect is considered together with harvesting and environmental noise. Stochastic description of population models extended with harvesting function provides motivation for both studies of population dynamics and studies that cross the boundaries of population dynamics. Even if choosing harvesting function linear in a stochastic population model showing Allee effect presents useful formulation to discuss effects of noise, this is not a sufficiently realistic choice. Therefore, in this article, we extend our analysis by specifying the harvesting as a commonly observed functional response in nature, Holling type-II. There are also many predator-prey models that include the Holling type-II functional response. \cite{Sen2019,Song2012,Peng2009,Zeng2008} \par

A simple population model showing Allee effect can be represented by a cubic model. Although deterministic models have pedagogical significance, they do not produce sufficiently realistic results. Although there are various approaches to describe the model including stochastic properties with noise, two approaches are basically in the foreground \cite{Yu2018a,Assaf2017b}. Master equation can be solved by using transition rates representing the transition of a population from $n$ individuals to $n+r$ individuals \cite{gardiner,Assaf2010a,Eskin2019}. Although there is no general solution of the master equation, a mean field equation can be obtained \cite{Mendez2019a}. Another way to obtain a stochastic model representing the population under internal and external fluctuations is to write a Fokker-Planck equation \cite{gardiner,jacobs}. For example, a Langevin equation is written using the statistical properties of Gaussian white noise and the Fokker-Planck equation is obtained from the correlation relations. In fact, the choice of the type of noise is important. There are studies that use colored noise to represent stochastic fluctuations \cite{Yang2019a,Levine2013a,Yang2018}. \par

White and colored noises effect populations differently, so the color of the noise is one of a primary things to consider while estimating populations probabilities of extinction \cite{Wilson1997,Kamenev2008,Ripa1996,Schwager2006,Johst1997}. While considering different types of populations subject to different environmental and intrinsic conditions, color of the noise should be chosen properly \cite{Cuddington1999,Steele1985}. In the case of white noise, values of noise at different times are independent of each other and autocorrelation is zero. But in the case of colored noise, noise is positively autocorrelated, meaning values of noise at contiguous times depend on each other and they are likely to be similar. For first time in this study, we consider the effect of correlated white and correlated colored noises on a population model showing Allee effect and subject to harvesting. It is important for noises to be correlated to model populations in a more realistic way. 
In this paper it is shown that, when a population is subject to stochastic fluctations, presence of harvesting causes weak Allee effect to turn to strong Allee effect and reduces probability of population to settle around the carrying capacity, a stable stationary point. This means that the population will extinct faster than expected. Controlling the size of a population showing Allee effect contributes both empirical and theoretical studies about determining optimum harvesting and determining an effective strategy in terms of conservation of ecological balance. Also, this study may contribute to determination of a conservation strategy for biological invasions. There are empirical studies reporting that the functional response of the predator against the changes in the population density of prey is a Holling type-II kind of functional response. For this reason, the case in which harvesting is expressed with a Holling type-II kind of function is also important for determining the dynamic evolution of the population by understanding how the populations showing Allee effect will be affected by stochastic fluctuations and to develop correct conservation strategies.\par

Paper is organized as follows. We expand the cubic deterministic model with harvesting function and carry a detailed study on the harvesting for both weak and strong Allee effect cases in section \ref{sec2}. Then, in section \ref{sec3}, we obtain the Stationary Probability Distribution (SPD) by making a stochastic description of a population model subject to white noise and harvesting effect with the Fokker-Planck equation. We discuss the effects of multiplicative noise strength, additive noise strength and the degree of correlation between additive and multiplicative noise on SPD for the population model. After that, we look at how linear and nonlinear harvesting (Holling type-II) effect SPD. In section \ref{sec4},  we consider population model subject to colored noise and we obtain SPD from the Approximate Fokker-Planck Equation. Then, we discuss the effect of cross-correlation time and degree of correlation between colored noises on SPD. After, we look at the effects of linear and nonlinear harvesting on SPD for population subject to colored noise. In section \ref{sec5} we conclude our results.

\section{Cubic Deterministic Model Under Allee Effect in the Presence of Harvesting} \label{sec2}

When the number of births per unit population is proportional to the square of the population size, and the Allee effect taken into consideration, a standard model is obtained where the time change of the population size is a cubic function of the population size \cite{Volterra1978}:

\begin{eqnarray} \label{eq:1}
\frac{{dx}}{{dt}} = rx\left( {1 - \frac{x}{K}} \right)\left( {x - m} \right)
\end{eqnarray}

Here, $x$  represents the average population size at time $t$, $r$ is the intrinsic growth rate, $K$ is the carrying capacity, and $m$ is the threshold value for population size below which population goes to extinction. For the model to be biologically meaningful, parameters other than $m$ must be positive. In the population model, $m>0$ should be used to describe the strong Allee effect, and $m<0$ to describe the weak Allee effect. In addition, $m<K$ because of the biological definitions of parameters. This model is deterministic in terms of mathematically and precisely describing the future state of the population when the initial conditions are known. By subtracting an $H(x)$ function, representing harvesting, from the right side of the equation (2.1) we expand the model as follows:

\begin{eqnarray} \label{eq:2}
\frac{{dx}}{{dt}} =  - \frac{r}{K}{x^3} + r\left( {1 + \frac{m}{K}} \right){x^2} - rmx - H(x)
\end{eqnarray}

When $H(x)=0$, it is clear that equation (2.2) takes the standard form in (2.1).
In Fig. 1, change of average population size over time for different initial population values is shown as carrying capacity $K=30$, Allee threshold $m=10$  and intrinsic growth rate $r=0.1$ taken for a population under strong Allee effect. Average population size always reaches the carrying capacity for different initial values of average population size above Allee threshold. In addition to this, for initial values below Allee threshold, average population size always reaches to zero. Thus, the deterministic model predicts that the average population size for each value above the Allee threshold will go to the carrying capacity, which is an equilibrium point, and will also go to another equilibrium point, zero, for each value below the Allee threshold. In Fig. 2, population under weak Allee effect shown. In this case, average population size, regardless of the initial value, except that it is 0,  always goes to the carrying capacity, which is an equilibrium point. Fig. 3 shows the relationship between the average population size and the derivative of average population size with respect to time. When $dx/dt=0$  taken, average population size values, which are the solution of equation (2.2) defined as equilibrium points. For strong Allee effect case, $x=0$ is stable, $x=m$ is unstable, and $x=K$ are stable equilibrium points. 

\begin{figure}[ht]
\includegraphics[scale=0.24]{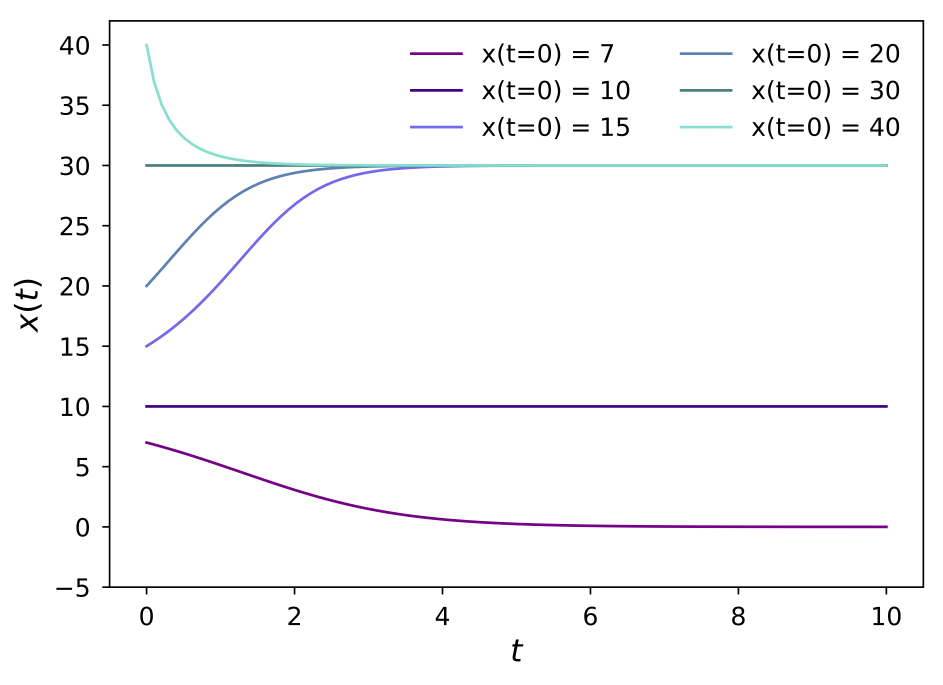}%
\caption{Average population size $x(t)$ as a function of $t$ for different initial $x$ values when, $K=30$, $r=0.1$, $m=10$.}
\end{figure}

\begin{figure}[ht]
\includegraphics[scale=0.24]{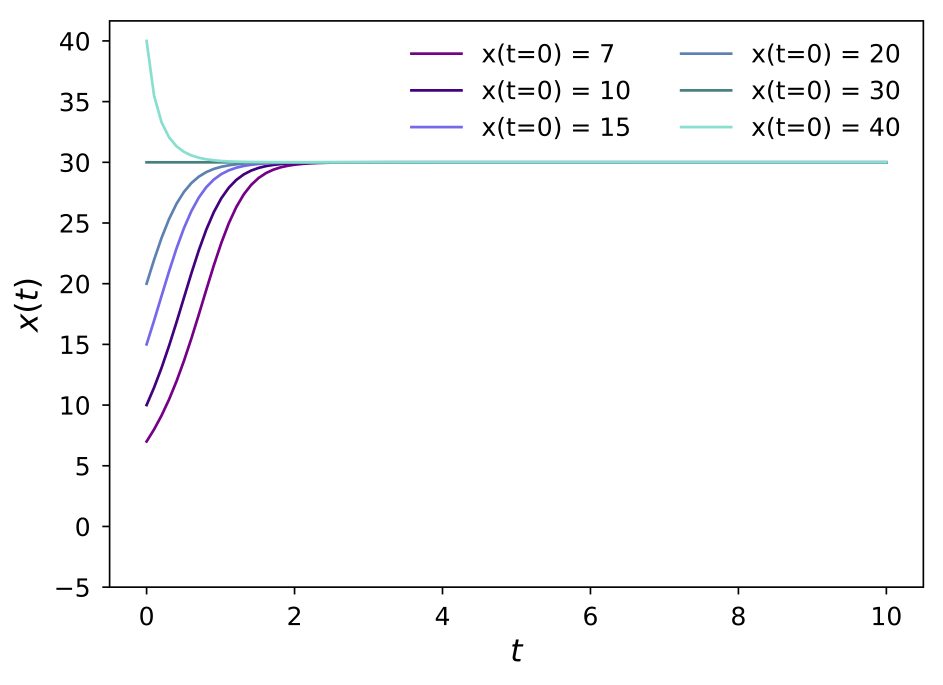}%
\caption{Average population size $x(t)$ as a function of $t$ for different initial $x$ values when, $K=30$, $r=0.1$, $m=-10$.}
\end{figure}

\begin{figure}[ht]
\includegraphics[scale=0.24]{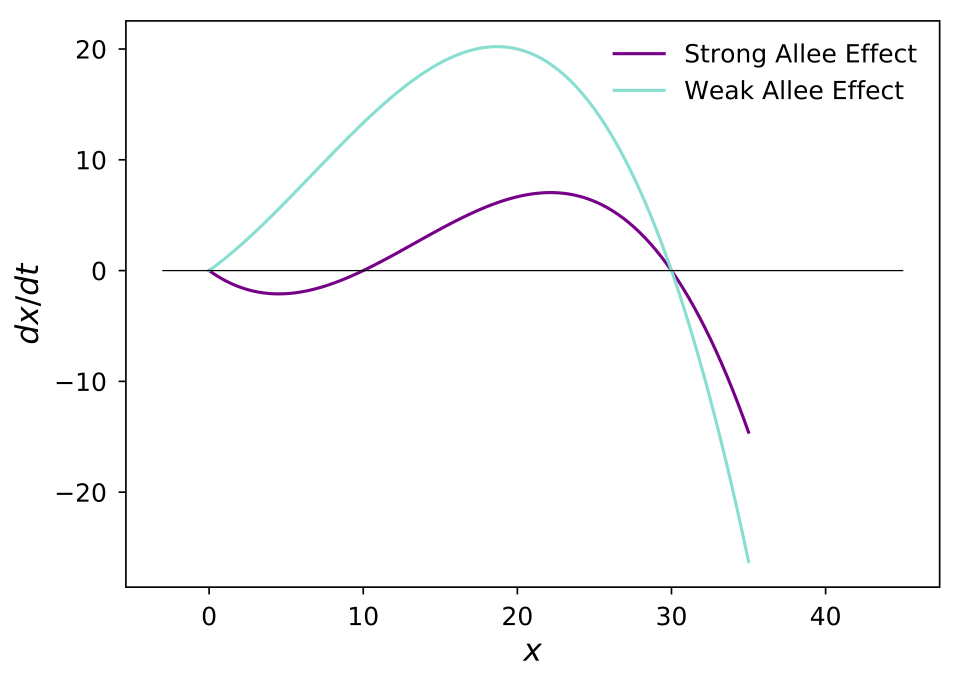}%
\caption{Time derivative of average population size $dx/dt$ as a function of $x$ when, $K=30$, $r=0.1$ and $m=10$, $m=-10$ for strong and weak Allee effects.}
\end{figure}

\begin{figure}[ht]
\includegraphics[scale=0.24]{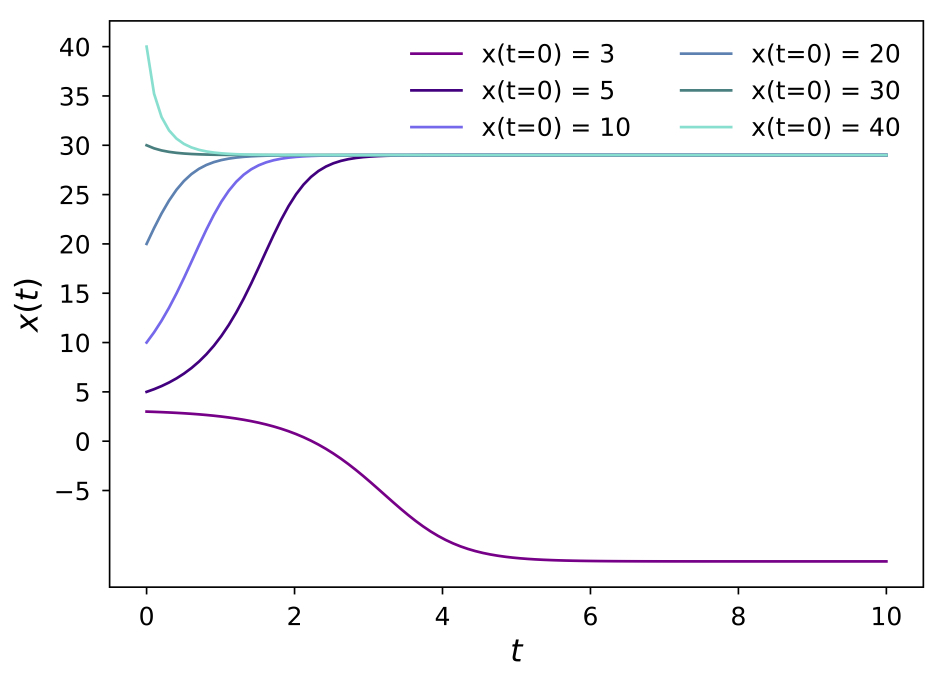}%
\caption{Average population size $x(t)$ as a function of $t$ for different initial $x$ values when, $K=30$, $r=0.1$, $m=-10$ and $H(x)=3.75$.}
\end{figure}

\begin{figure}[ht]
\includegraphics[scale=0.24]{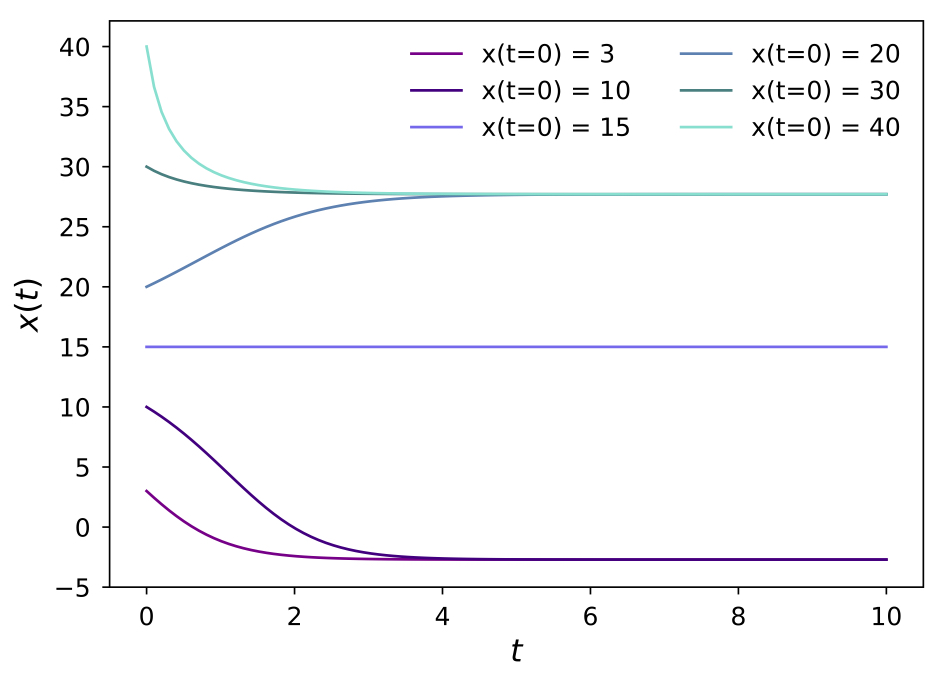}%
\caption{Average population size $x(t)$ as a function of $t$ for different initial $x$ values when, $K=30$, $r=0.1$, $m=10$ and $H(x)=3.75$.}
\end{figure}

\begin{figure}[ht]
\includegraphics[scale=0.24]{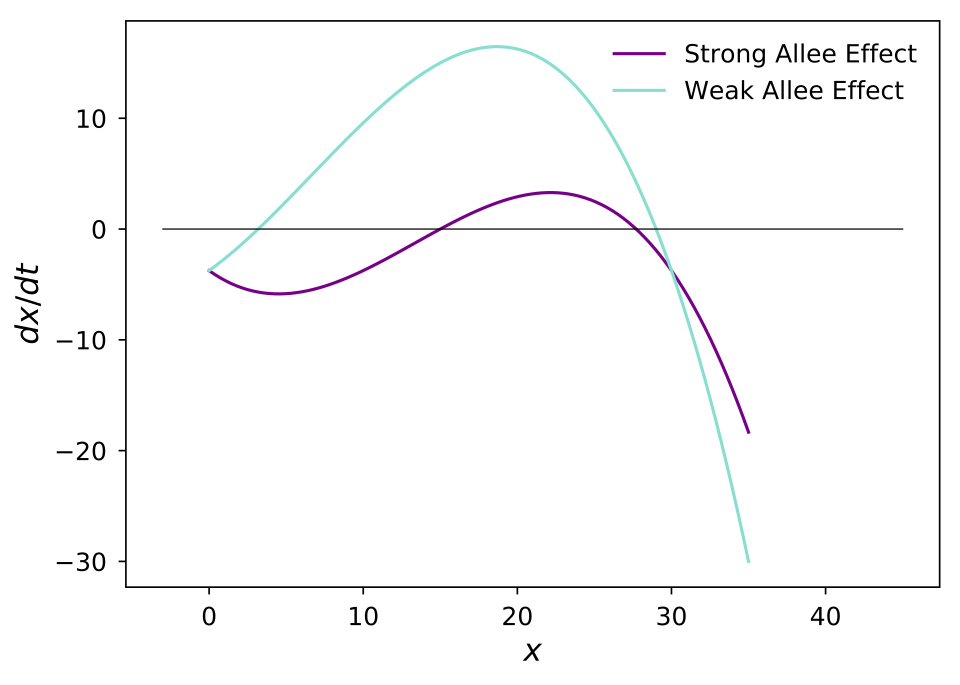}%
\caption{Time derivative of average population size $dx/dt$ as a function of $x$ when, $K=30$, $r=0.1$, $H(x)=3.75$, and $m=10$, $m=-10$ for strong and weak Allee effects.}
\end{figure}

\begin{figure}[ht]
\includegraphics[scale=0.24]{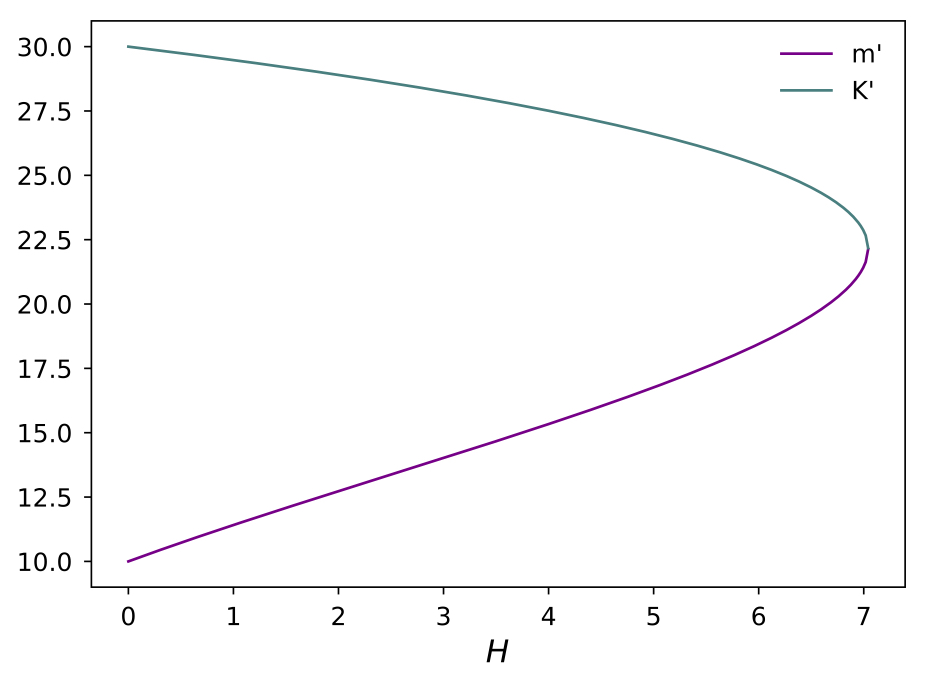}%
\caption{Change of Allee threshold, $m^{'}$, and carrying capacity, $K^{'}$ with respect to $H$, when $K=30$, $r=0.1$, $m=10$.}
\end{figure}
In this case, it is clearly seen in Fig. 3 that when $0<x<m$, $dx/dt$  becomes negative and average population size goes to zero but when  $m<x<K$, $dx/dt$ becomes positive so that average population size goes to carrying capacity. For the weak Allee effect case, there are only two non-negative roots, $x=0$  and $x=K$. \par
We continue by examining the situation where the harvesting function is represented by a constant $H(x)=H$ and $H>0$. An interesting result here is that in the presence of a harvesting term independent of the average population size, weak Allee effect can no longer be mentioned. The presence of the harvesting function as a constant in the model does displace not only the roots but also causes the formation of a root above zero for the weak Allee effect case, that is, an Allee threshold that did not exist before. This case is clearly seen in Fig. 4.
There is now such a critical average population size value just above zero, where the average population size is going to zero at each initial average population size value below this value and goes to carrying capacity for each initial value above this threshold. So, the model now foresees an Allee threshold. As the harvesting term grows, the value of the Allee threshold increases and the carrying capacity of the population decreases. The steady decline of the average size of \mbox{population} means an external environmental condition that suppresses the carrying capacity of the population. As can be seen from Fig. 5, in the presence of harvesting term, the model that was showing Allee effect before the addition of harvesting term, has $x(t)=15$ as Allee threshold even though $m=10$. In other words, in the model (2.2), where harvesting term is 0, $m$ stands as the Allee threshold. However, expanding the model to include a fixed harvesting term causes displacement of the equilibrium points. Thus, $m$ no longer represents the Allee threshold of the model. Also, $K$, which was the carrying capacity of the population before the addition of harvesting term, is no longer carrying capacity. In the presence of a constant harvesting term carrying capacity of the population has dropped below $K$. 
Fig. 6 shows the change of roots in the presence of constant harvesting term clearly. From now on even if the $x=0$ is not a root when $dx/dt=0$ taken, it is biologically meaningful. Because, near $x=0$, $dx/dt<0$, so that the average population size value must reduce. However, $x=0$ acts as a biologically stable equilibrium since the average population size cannot be less than zero biologically. Therefore, mathematically, in Figures 4 and 5, the fact that the average population size is less than zero does not have a biological significance and is biologically interpreted as the reset of the average population size. Fig. 7 shows the change of the carrying capacity of the population and Allee threshold according to the harvesting.
We discussed the change of equilibrium points for the weak and strong Allee effect cases with the expansion of the population model. To determine equilibrium points, we can find positive roots by taking $dx/dt=0$ again in equation (2.2). Thus, after expanding the model to include a fixed harvesting term, we obtained the following solutions for the new Allee threshold $m^{'}$ and the new carrying capacity $K^{'}$ as:

\begin{eqnarray}
m' = A\sin \left( {\frac{1}{3}\arcsin E} \right) + \frac{{K + m}}{3}\\
K' = A\cos \left( {\frac{\pi }{6} + \frac{1}{3}\arcsin E} \right) + \frac{{K + m}}{3}
\end{eqnarray}

Where $E$ and $A$ given as:

\begin{subequations}
\begin{eqnarray}
E =&& \frac{{3\sqrt 3 }}{2}\left[ { - \frac{2}{{27}}{{(K + m)}^3} + \frac{1}{3}mK(K + m) +\frac{{KH}}{r}} \right]\nonumber \\
&&\times{\left[ {\frac{{{{(K + m)}^2}}}{3} - mK} \right]^{ - 3/2}}
\end{eqnarray}
\begin{eqnarray}
A = \frac{2}{{\sqrt 3 }}{\left[ {\frac{{{{(K + m)}^2}}}{3} - mK} \right]^{1/2}}
\end{eqnarray}
\end{subequations}

We continue by assuming harvesting function to be in the form of $H(x)=H_1x^2 +H_2x$ . When $dx/dt=0$ taken and $r^2(1+m/K-H_1/r)^2>4r(rm+H_2)/K$ , equation (2.2) arranged as follows and $m^{'}$, $K^{'}$ are the roots.

\begin{eqnarray}
\frac{{dx}}{{dt}} = \left( {rm + {H_2}} \right)x\left( {1 - \frac{x}{{K'}}} \right)\left( {\frac{x}{{m'}} - 1} \right)
\end{eqnarray}

Here $K^{'}$ and $m^{'}$ given as:
\begin{widetext}
\begin{eqnarray}
K' = \frac{K}{{2r}}\left[ {r\left( {1 + \frac{m}{K} - \frac{{{H_1}}}{r}} \right) + \sqrt {{r^2}{{\left( {1 + \frac{m}{K} - \frac{{{H_1}}}{r}} \right)}^2} - 4\frac{r}{K}\left( {rm + {H_2}} \right)} } \right]\\
m' =\frac{K}{{2r}}\left[ {r\left( {1 + \frac{m}{K} - \frac{{{H_1}}}{r}} \right) - \sqrt {{r^2}{{\left( {1 + \frac{m}{K} - \frac{{{H_1}}}{r}} \right)}^2} - 4\frac{r}{K}\left( {rm + {H_2}} \right)} } \right]
\end{eqnarray}
\end{widetext}

\begin{figure}[ht]
\includegraphics[scale=0.24]{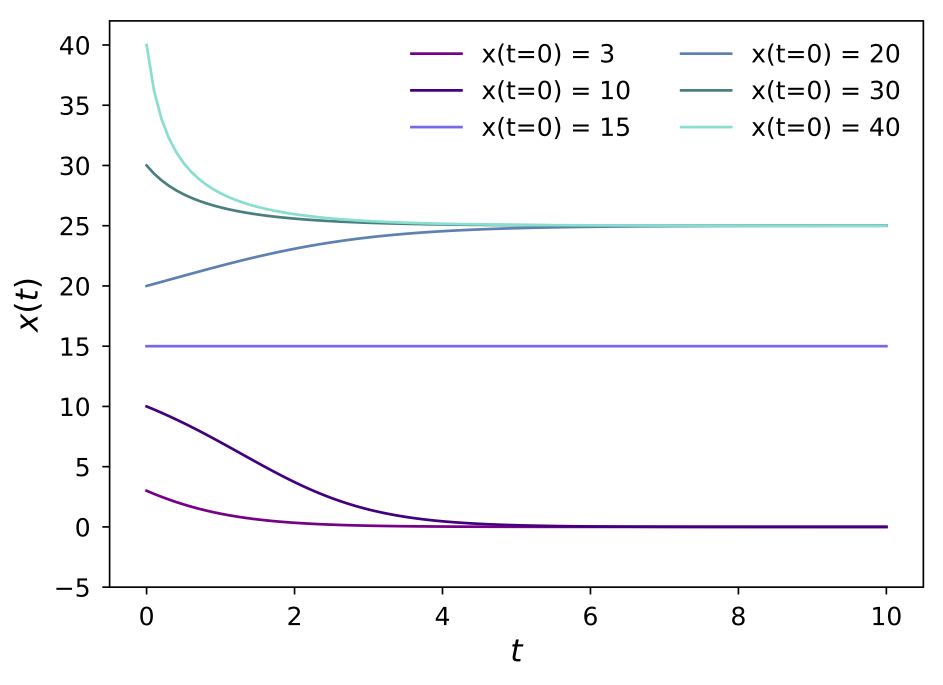}%
\caption{Average population size $x(t)$ as a function of $t$ for different initial $x$ values when, $K=30$, $r=0.1$, $m=10$ and $H(x)=0.25x$.}
\end{figure}

\begin{figure}[h]
\includegraphics[scale=0.24]{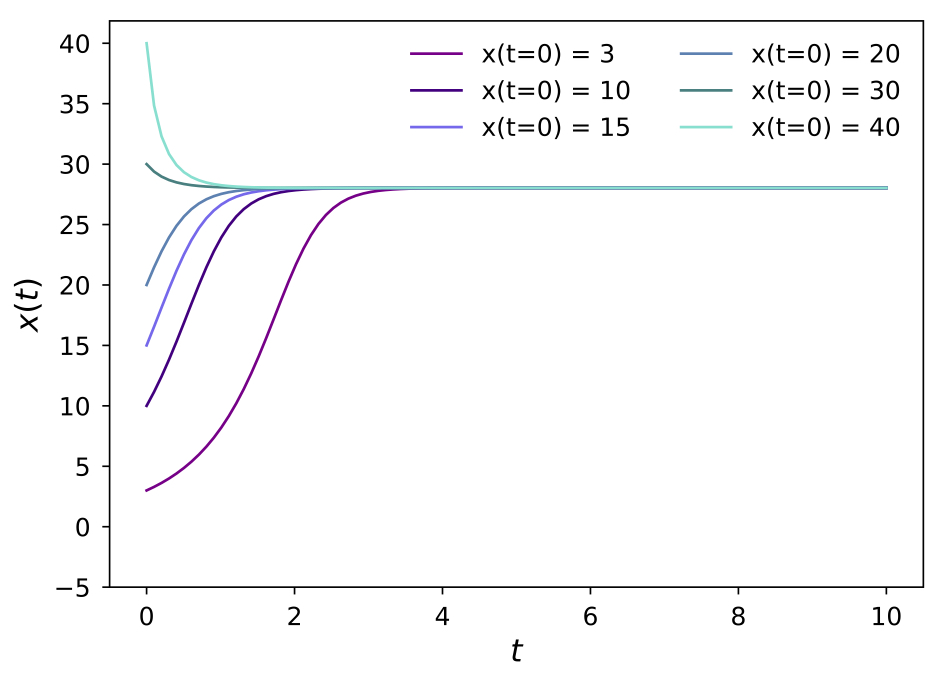}%
\caption{Average population size $x(t)$ as a function of $t$ for different initial $x$ values when, $K=30$, $r=0.1$, $m=-10$ and $H(x)=0.25x$.}
\end{figure}

When $H_1=0$, harvesting term is a linear function of average population size. In this case, new Allee threshold and carrying capacity can be calculated by (2.7) and (2.8). Furthermore, an advantage of writing (2.2) in the form of (2.6) is that the model represents either the weak and strong Allee effect according to the sign of  $rm+H_2$ expression. If $rm+H_2>0$ , model represents strong Allee effect, and if $rm+H_2<0$ represents weak Allee effect. When $H_1=0$ and, $H_2=0.25$ Fig. 8 stands for strong Allee effect, Fig. 9 for weak Allee effect, and Fig. 10 for relationship between the first derivative of average population size with respect to time and average population size.
In Fig. 9, it is seen that the harvesting term has the effect of decreasing the carrying capacity of the population under weak Allee effect. In addition to this, in Figure 8 it is seen that the harvesting term has an effect of decreasing the carrying capacity and increasing the Allee threshold in the population under strong Allee effect. An important difference between the case where the harvesting term is a linear function of the average size of the population and the case where the harvesting term is taken as a constant independent of the average size of the population is that the weak Allee effect does not immediately convert to a strong Allee effect in the first case. This is clear when Fig. 10 is examined. In Fig. 11, change of carrying capacity of the population and Allee threshold with respect to amount of harvesting shown. Figures 12 and 13 show the variation of the Allee threshold and the carrying capacity of the population depending on the coefficients $H_1$ and $H_2$ for population model under strong Allee effect where $H_1> 0$ and $H_2> 0$.

\begin{figure}[ht]
\includegraphics[scale=0.24]{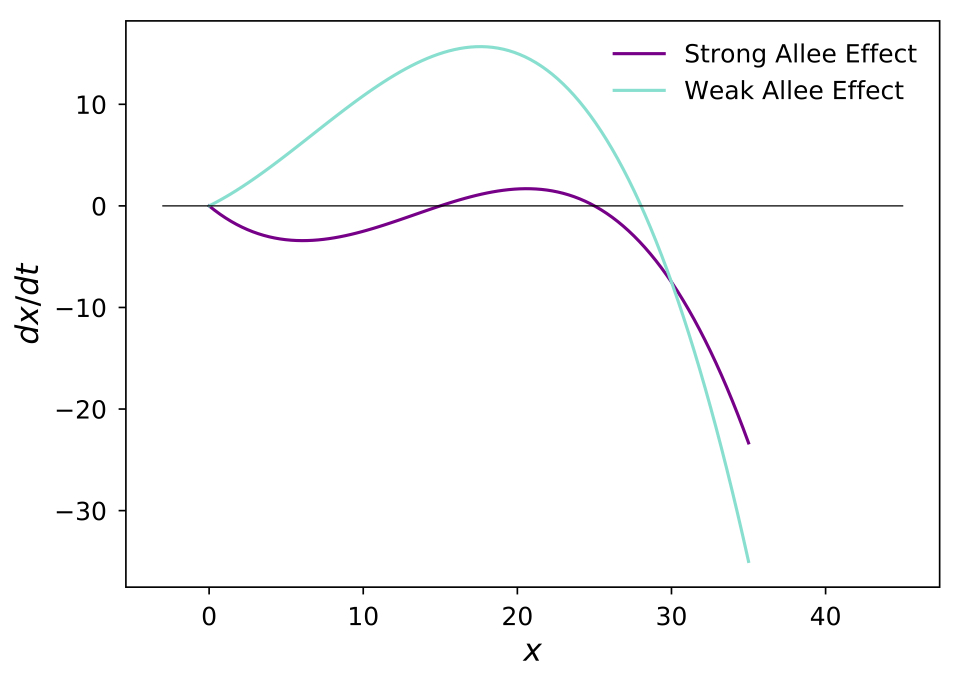}%
\caption{Time derivative of average population size $dx/dt$ as a function of $x$ when, $K=30$, $r=0.1$, $H(x)=0.25x$, and $m=-10$ for strong and weak Allee effect cases respectively.}
\end{figure}

\begin{figure}[ht]
\includegraphics[scale=0.24]{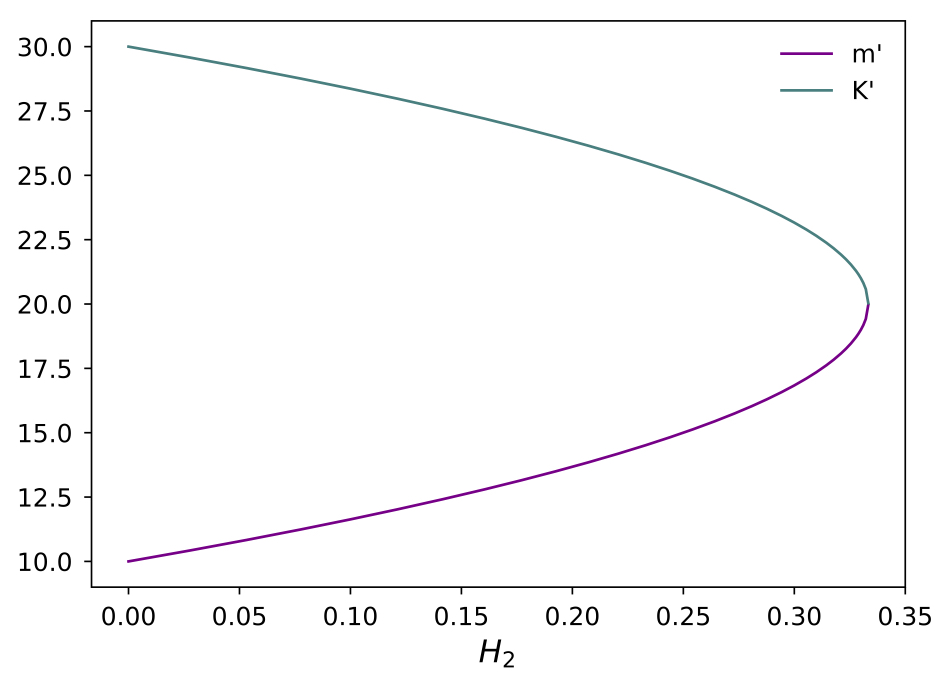}%
\caption{Change of Allee threshold, $m^{'}$ and carrying capacity, $K^{'}$ with respect to $H_2$, when $K=30$, $r=0.1$, $m=10$.}
\end{figure}

\begin{figure}[ht]
\includegraphics[scale=0.24]{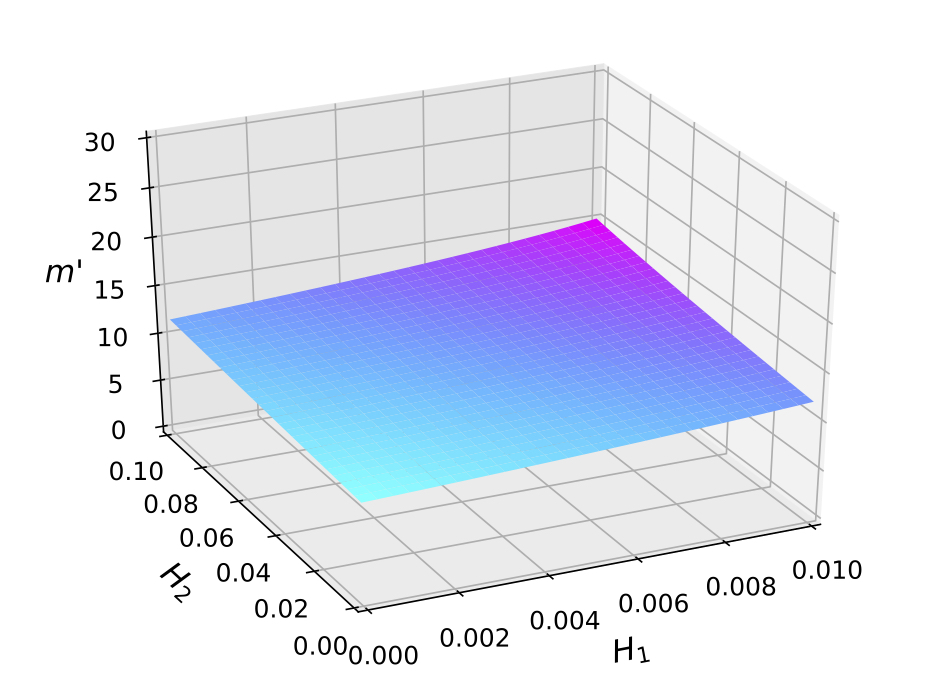}%
\caption{Change of Allee threshold, $m^{'}$, with respect to $H_1$ and $H_1$, when\\ $K=30$, $r=0.1$, $m=10$.}
\end{figure}

\begin{figure}[ht]
\includegraphics[scale=0.24]{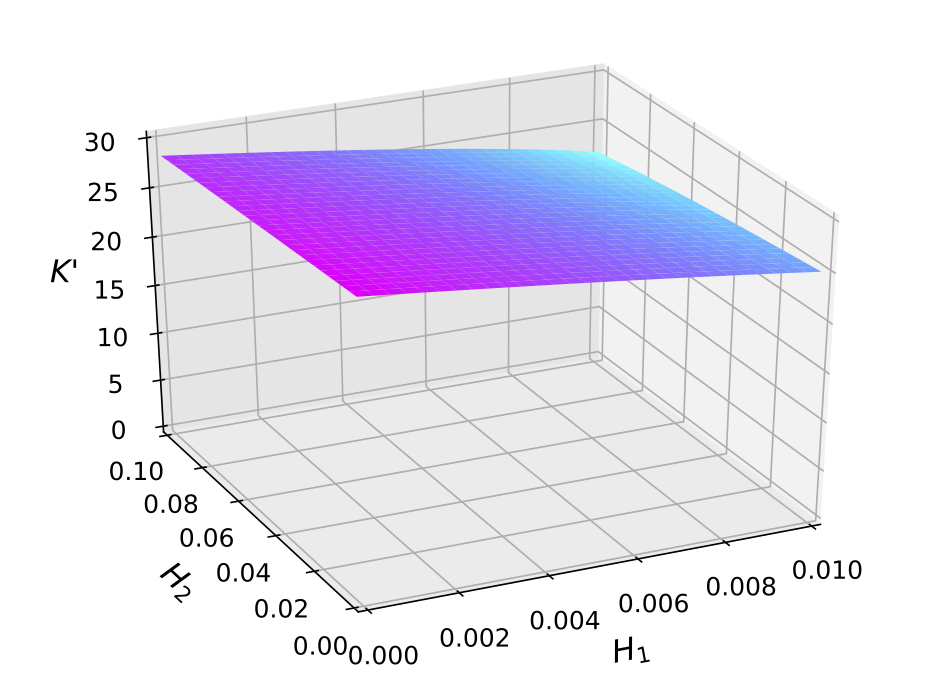}%
\caption{Change of carrying capacity, $K^{'}$ with respect to $H_1$ and $H_2$, when $K=30$, $r=0.1$,\\ $m=10$.}
\end{figure}

\newpage
\section{Stationary Probability Distribution of Population Subject to White Noise and Harvesting} \label{sec3}

In order to describe the cubic population model containing the Allee effect, which we discussed in the previous section, by stochastic approach, we write a Langevin equation as follows:

\begin{eqnarray}
\frac{{dx}}{{dt}} =&&  - \frac{r}{K}{x^3} + r\left( {1 + \frac{m}{K}} \right){x^2} \nonumber\\
&&- rmx -H(x) + x\zeta (t) + \psi (t)
\end{eqnarray}
Statistical properties of $\zeta (t)$ and $\psi (t)$, which represent Gaussian white noise with zero mean, defined as:

\begin{subequations}
\begin{align}
&\left\langle {\zeta (t)} \right\rangle  = \left\langle {\psi (t)} \right\rangle  = 0\\
&\left\langle {\zeta (t)\zeta (t')} \right\rangle  = 2D\delta (t - t')\\
&\left\langle {\psi (t)\psi (t')} \right\rangle  = 2\alpha \delta (t - t')\\
&\left\langle {\zeta (t)\psi (t')} \right\rangle  = 2\lambda \sqrt {D\alpha } \delta (t - t') \\
&\left\langle {\zeta (t)\psi (t')} \right\rangle  = \left\langle {\psi (t)\zeta (t')} \right\rangle
\end{align}
\end{subequations}

Here, parameters $D$ and $\alpha$ are the strength of the noises $\zeta(t)$ and $\psi(t)$ and the parameter $\lambda$ is the degree of correlation between $\zeta(t)$ and $\psi(t)$. By using (3.1) and (3.2) and by following the general procedure given in \cite{Da-Jin1994}, we arrive at a Fokker Planck equation:

\begin{eqnarray}
\frac{\partial }{{\partial t}}P(x,t) =&&  - \frac{\partial }{{\partial x}}\left[ {A(x)P(x,t)} \right] \nonumber \\
&&+ \frac{{{\partial ^2}}}{{\partial {x^2}}}\left[ {B(x)P(x,t)} \right]
\end{eqnarray}

\begin{subequations}
\begin{eqnarray}
A(x) = && - \frac{r}{K}{x^3} + r\left( {1 + \frac{m}{K}} \right){x^2} - rmx -H(x) \nonumber \\
&&+ Dx + \lambda \sqrt {D\alpha } 
\end{eqnarray}
\begin{eqnarray}
B(x) = D{x^2} + 2\lambda \sqrt {D\alpha } x + \alpha 
\end{eqnarray}
\end{subequations}

\bigskip
Stationary solution of Fokker Planck equation obtained by a simple conversion as follows:
\begin{eqnarray}
{P_{st}} = \frac{N}{{B(x)}}\exp \left[ {\int_{}^x {\frac{{A(x')}}{{B(x')}}dx'} } \right]
\end{eqnarray}

Here, $N$ is normalization constant. 

\subsection{Linear Harvesting}
When we take $H(x)=hx$, harvesting becomes linear function of $x$. Using equations (3.4) in (3.5), we find stationary probability distribution function $P_{st}$ for $0 \le \lambda  < 1$ as follows:

\begin{eqnarray}
{P_{st}} = \frac{{N{e^\eta }}}{{\sqrt {D{x^2} + 2\lambda \sqrt {D\alpha } x + \alpha } }}
\end{eqnarray}

Where,

\begin{widetext}
\begin{eqnarray}
\eta  = &&\frac{1}{{2K{D^3}}}\left[ {\varphi \ln \alpha  + 2\arctan \left( {\frac{{\lambda \sqrt {\alpha D} }}{{\sqrt {\alpha D - {\lambda ^2}\alpha D} }}} \right)\gamma  + 2Drx\left( {D(K + m) + 2\lambda \sqrt {\alpha D} } \right)\varphi } \right.\nonumber\\
 &&\left. { - {D^2}r{x^2}\varphi  - \varphi \ln \left( {\alpha  + D{x^2} + 2\lambda x\sqrt {\alpha D} } \right) - 2\arctan \left( {\frac{{Dx + \lambda \sqrt {\alpha D} }}{{\sqrt {\alpha D - {\lambda ^2}\alpha D} }}} \right)\gamma } \right]\\
\varphi  =&& \left[ {D( - \alpha r + DK(h + mr)) + 2D\lambda (K + m)r\sqrt {\alpha D}  + 4{\lambda ^2}r\alpha D} \right]\\
\gamma  =&& \frac{{\left( {\alpha {D^2}(K + m)r - D\lambda \left( { - 3\alpha r + DK(h + mr)} \right)\sqrt {\alpha D}  - 2D{\lambda ^2}(K + m)r\alpha D - 4{\lambda ^3}r{(\sqrt{\alpha D})^{{\raise0.3ex\hbox{$3$}}}}} \right)}}{{\sqrt {\alpha D - {\lambda ^2}\alpha D} }}
\end{eqnarray}
\end{widetext}

Thus, for a population model under Allee effect, in the presence of Gaussian white noise and in the presence of linear harvesting function, the stationary probability distribution function of the average size of  population was obtained from the solution of the Fokker-Planck equation.

\begin{figure}[h]
\includegraphics[scale=0.24]{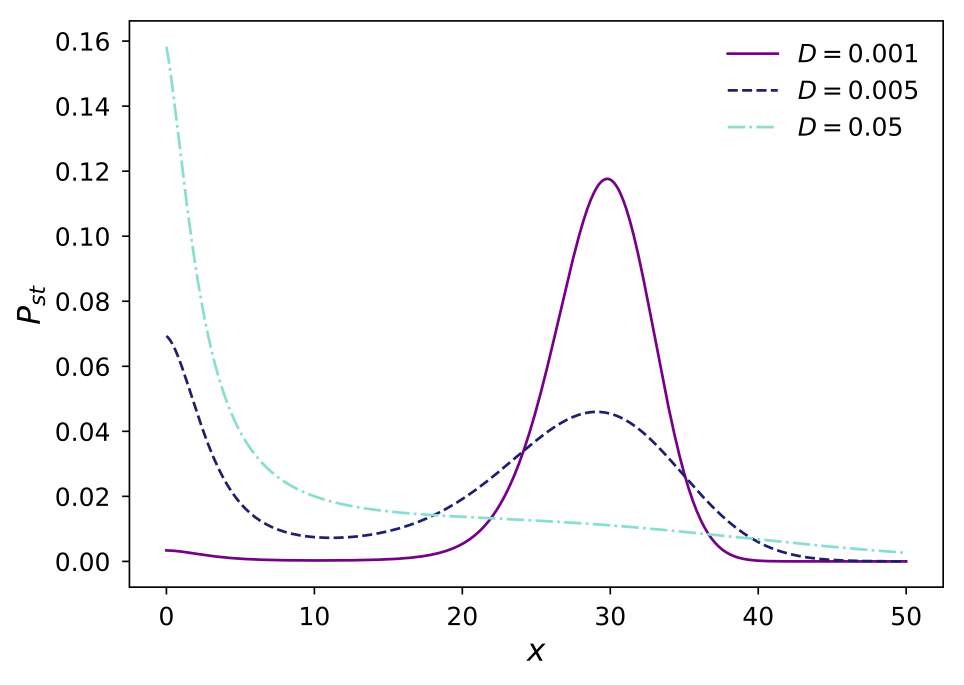}%
\caption{$P_{st}$ as a function of $x$ for different values of $D$ when $r=0.01$, $K=30$, $m=10$, $\alpha=0.5$ and $\lambda=0.5$}
\end{figure}

\begin{figure}[ht]
\includegraphics[scale=0.24]{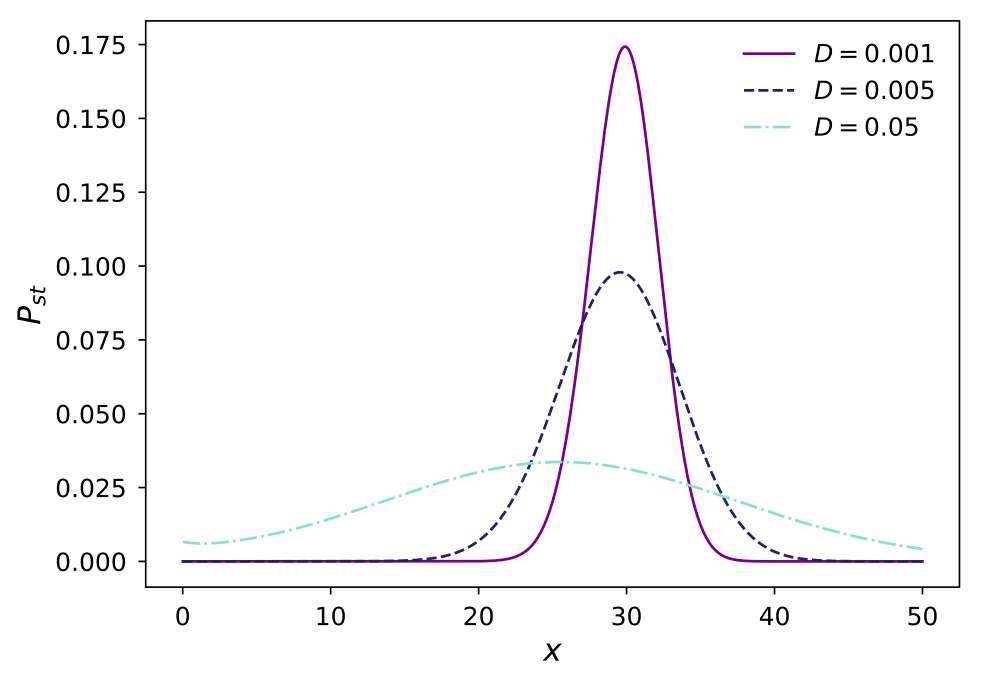}%
\caption{$P_{st}$ as a function of $x$ for different values of $D$ when $r=0.01$, $K=30$, $m=-10$, $\alpha=0.5$ and \\ $\lambda=0.5$.}
\end{figure}

First, we discuss the effects of additive and multiplicative noise for the case there is no harvesting.
In Fig. 14, change of stationary probability distribution function (SPD) for different values of strength of multiplicative noise $D$ given for population under strong Allee effect, where other parameters are $r=0.01$, $K=30$, $m=10$, $\alpha=0.5$ and $\lambda=0.5$. For small values of the strength of multiplicative noise, when population size is near the carrying capacity, SPD has a one peak. In other words, for the very small values of $D$, the population is likely to be around the carrying capacity. With the increase of $D$, probability of population to take values farther away from the carrying capacity gets bigger by the spread of SPD around carrying capacity. In further growing values of $D$, the probability of the population to be near carrying capacity decreases and the probability to be at low population size values increases. Therefore, we say that the high strength of multiplicative noise under strong Allee effect will increase the chance of population to extinct. \par
In Fig. 15 change of SPD for different values of the strength of multiplicative noise for population under weak Allee effect given. For that case, $m =-10$ and other parameters are same as the previous case. For very small values of multiplicative noise, we also see a peak in carrying capacity but for that case probability of population to be at carrying capacity is higher compared to the strong Allee effect case. Increasing the strength of multiplicative noise causes the SPD to become more widespread around carrying capacity. For large values of the strength of multiplicative noise, population does not extinct and SPD gets wider in shape.

\begin{figure}[ht]
\includegraphics[scale=0.24]{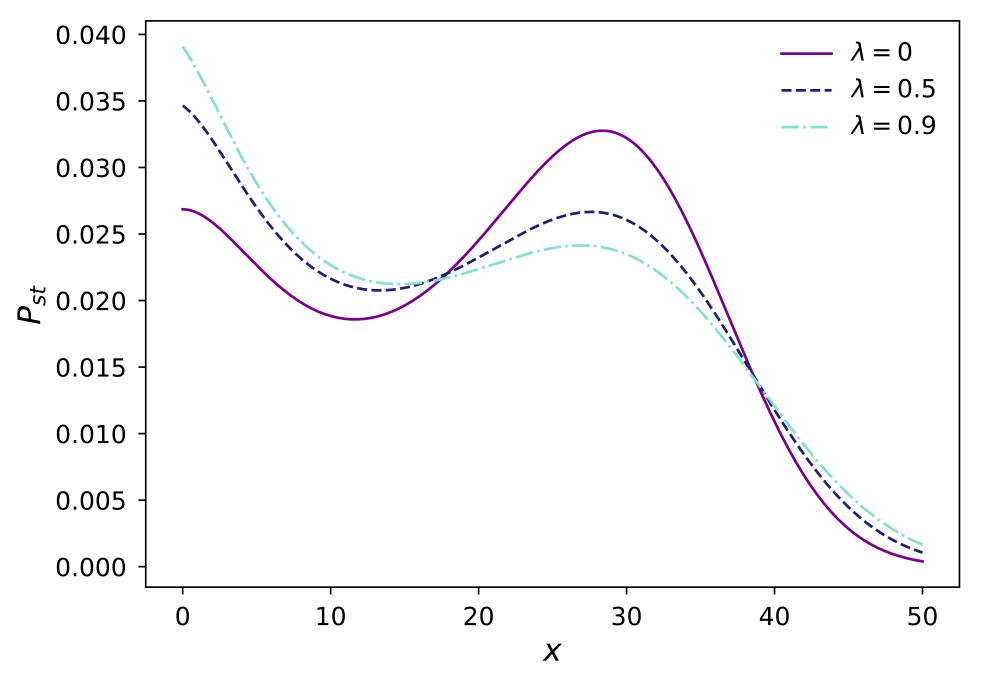}%
\caption{$P_{st}$ as a function of $x$ for different values of $\lambda$ when $r=0.01$, $K=30$, $m=10$, $\alpha=0.5$ and $D=0.001$.}
\end{figure}

\begin{figure}[ht]
\includegraphics[scale=0.24]{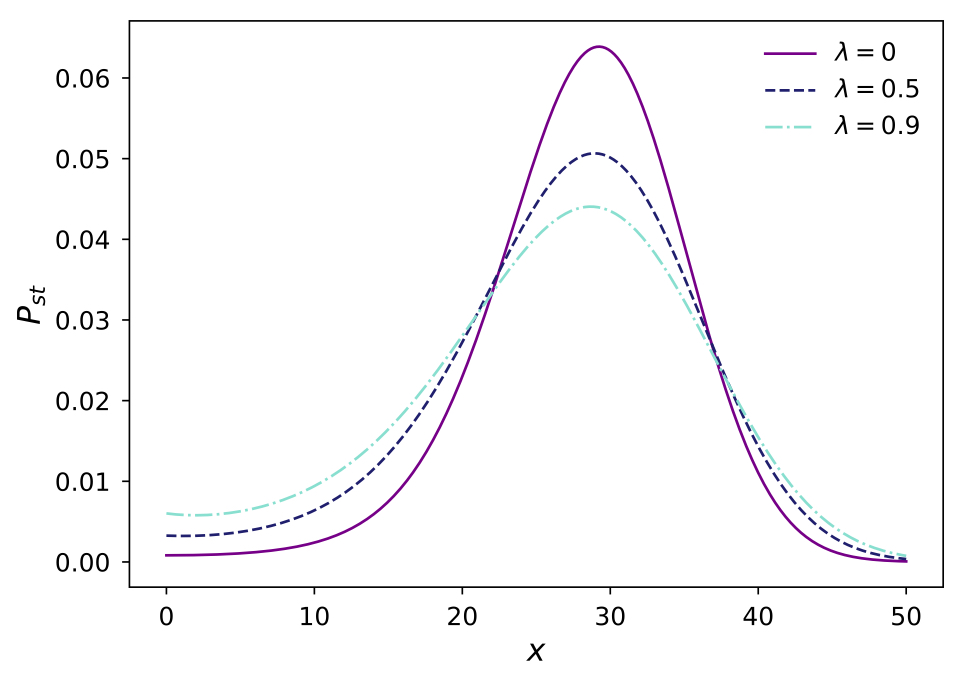}%
\caption{$P_{st}$ as a function of $x$ for different values of $\lambda$ when $r=0.01$, $K=30$, $m=-10$, $\alpha=0.5$ and $D=0.001$.}
\end{figure}

\par
In Fig. 16, where $r=0.001$, $K=30$, $m=10$, $D=0.001$ and $\alpha=0.5$, change of SPD for \mbox{different} values of the degree of correlation between noises given. In the presence correlation between noises, it is seen in Fig. 16 that the probability of the population size to be around the carrying capacity decreases and as the degree of correlation increases the probability of the population to be around $0$ increases. In Fig. 17, $m = -10$, but other parameters are the same as in the previous case. It shows that the presence of correlation between noises  in the case of a weak Allee effect reduces the likelihood of being around the carrying capacity of the population and increases the likelihood of being around 0, similar to the case of a strong Allee effect. From this, we say that increasing degree of correlation between noises increases tendency of \mbox{population} to extinct.

\begin{figure}[ht]
\includegraphics[scale=0.24]{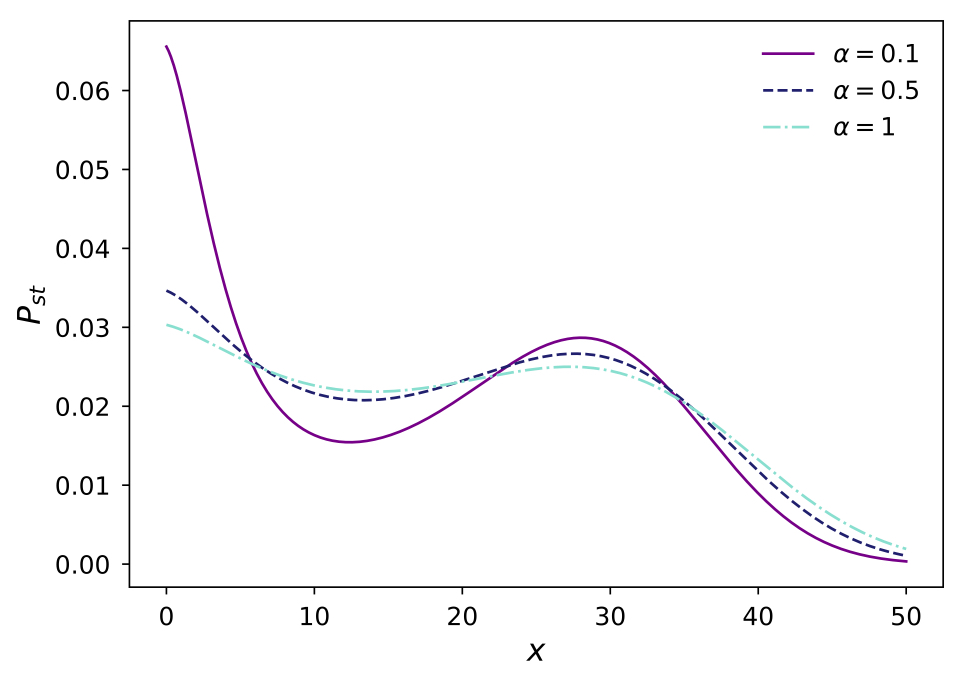}
\caption{$P_{st}$ as a function of $x$ for different values of $\alpha$ when $r=0.001$, $K=30$, $m=10$, $\lambda = 0.5$ \\and $D=0.001$.}
\end{figure}

\begin{figure}[ht]
\includegraphics[scale=0.24]{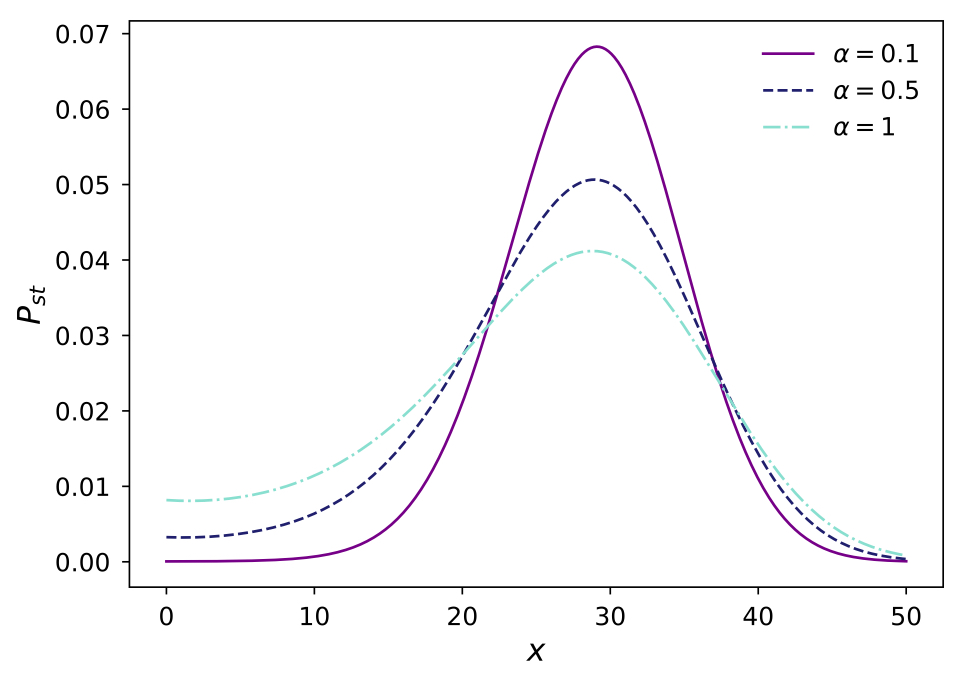}
\caption{$P_{st}$ as a function of $x$ for different values of $\alpha$ when $r=0.001$, $K=30$, $m=-10$, $\lambda = 0.5$ \\and $D=0.001$.}
\end{figure}

\begin{figure}[ht]
\includegraphics[scale=0.24]{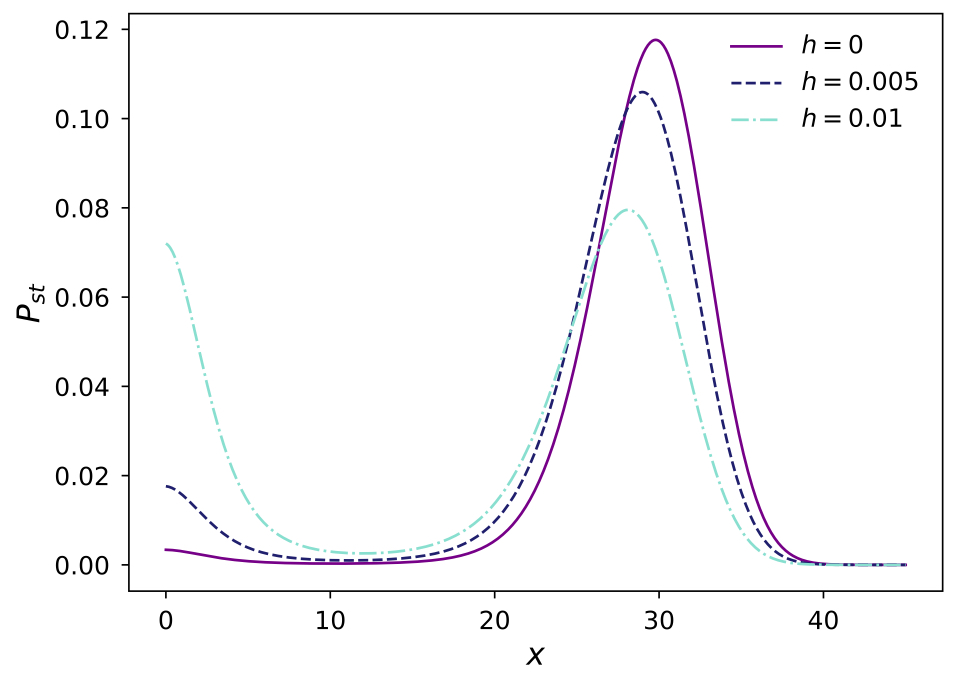}%
\caption{$P_{st}$ as a function of $x$ for different values of $h$ when $r=0.01$, $K= 30$, $m=10$, $D=0.001$, $\lambda=0.5$ \\ and $\alpha= 0.5$.}
\end{figure}

\begin{figure}[ht]
\includegraphics[scale=0.24]{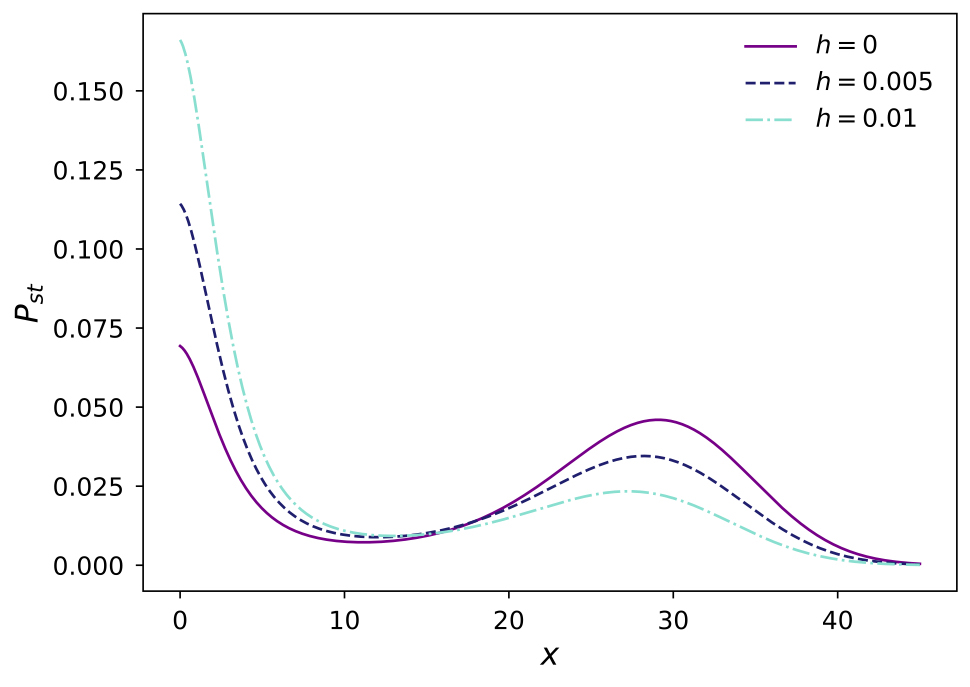}%
\caption{$P_{st}$ as a function of $x$ for different values of $h$ when $r=0.01$, $K= 30$, $m=10$, $D=0.005$, $\lambda=0.5$ and $\alpha= 0.5$.}
\end{figure}

\begin{figure}[ht]
\includegraphics[scale=0.24]{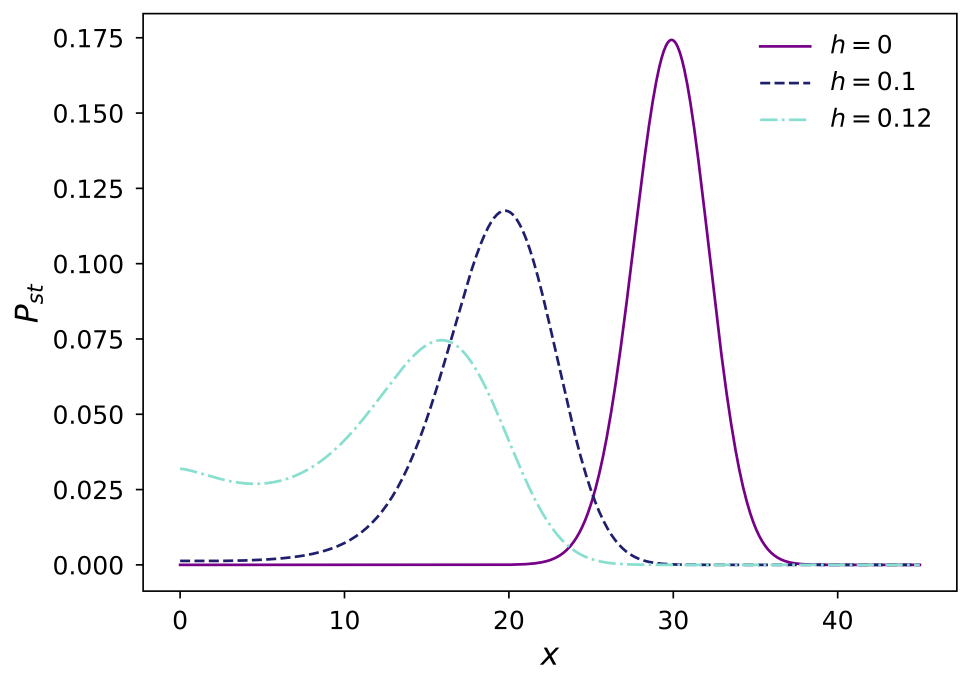}%
\caption{$P_{st}$ as a function of $x$ for different values of $h$ when $r=0.01$, $K= 30$, $m=-10$, $D=0.001$, $\lambda=0.5$ and $\alpha= 0.5$.}
\end{figure}

\begin{figure}[ht]
\includegraphics[scale=0.24]{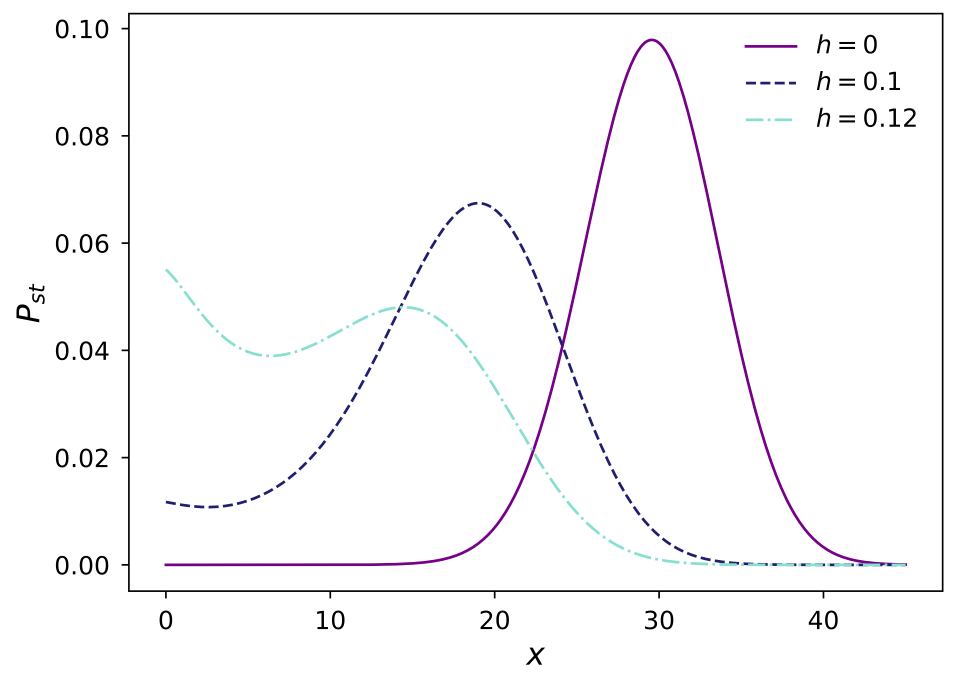}%
\caption{$P_{st}$ as a function of $x$ for different values of $h$ when $r=0.01$, $K= 30$, $m=-10$, $D=0.005$, $\lambda=0.5$ and $\alpha= 0.5$.}
\end{figure}

\par
In Fig. 18, the increase of the strength of additive noise under strong Allee effect, with $r = 0.001$, $K = 30$, $m = 10$, $D = 0.001$ and $\lambda = 0.5$, reduces the likelihood that the population size will be around 0 and carrying capacity, causes SPD to have more flattened shape. In addition to that, the reduction in the likelihood of being around the carrying capacity of the population is smaller compared to the reduction in the probability of being at 0. In this case, we can say that the increase in the strength of additive noise has a positive effect on the survival of the population in case of strong Allee effect. In Fig. 19, $m = -10$ and other \mbox{parameters} are taken same as the previous case. In the case of weak Allee effect, the increase in additive noise strength decreases the probability of the population size to be around the carrying capacity and increases the probability of being around 0. We say that in the case of the weak Allee effect, the increase in the intensity of additive noise increases the tendency of the population to extinct.
\par
In Fig. 20, for parameters $r=0.01$, $K= 30$, $m=10$, $D=0.001$, $\lambda=0.5$ and $\alpha= 0.5$, and in Fig. 21 for same parameters except that $D=0.005$, change of stationary probability distribution function (SPD) for different values of $h$ given for population under strong Allee effect. In the deterministic model we \mbox{examined} in Part II, it was discussed that the harvesting term has an effect that decreases the carrying capacity and increases the Allee threshold in the population under strong Allee effect. Figures 20 and 21 clearly show the change in carrying capacity. The growing values of $h$ cause a decrease in carrying capacity and a flattening of the SPD. Thus, for a population under a strong Allee effect, the effect of $h$ is that it decreases the probability of the population size to be around the carrying capacity and increases the probability of being close to zero. At the same time, we see that the strength of multiplicative noise, $D$, being greater causes the change in $h$ to have a stronger effect.

Figures 22 and 23 show the effect of $h$ in a \mbox{population} under weak Allee effect for $m = -10$, with the same parameters as the case of strong Allee effect except $m$. In the case of small $h$, SPD shows bimodal structure and in the case of bigger $h$, the shape of SPD turns to unimodal. The second peak occurs at population values close to zero. Bimodal structure of SPD shows that population will extinct under a threshold value. 

\subsection{Nonlinear Harvesting (Holling type-II)}
In this case, we consider harvesting function as a nonlinear function of $x$ in the form of $H(x)=hx/(b_1+b_2x)$. In Fig. 24, change of SPD with respect to $h$ for strong Allee effect case given. As $h$ increases probability of population to extinct increases whereas probability to be around carrying capacity decreases. Also, change of carrying capacity and Allee threshold is seen clearly. In Fig. 25, change of SPD for different values of $h$ given for weak Allee effect case. As $h$ values increase, SPD starts to show bimodal shape. Also, weak Allee effect turns to strong Allee effect, and in the same time probability of population to extinct increases.

\begin{figure}[ht]
\includegraphics[scale=0.24]{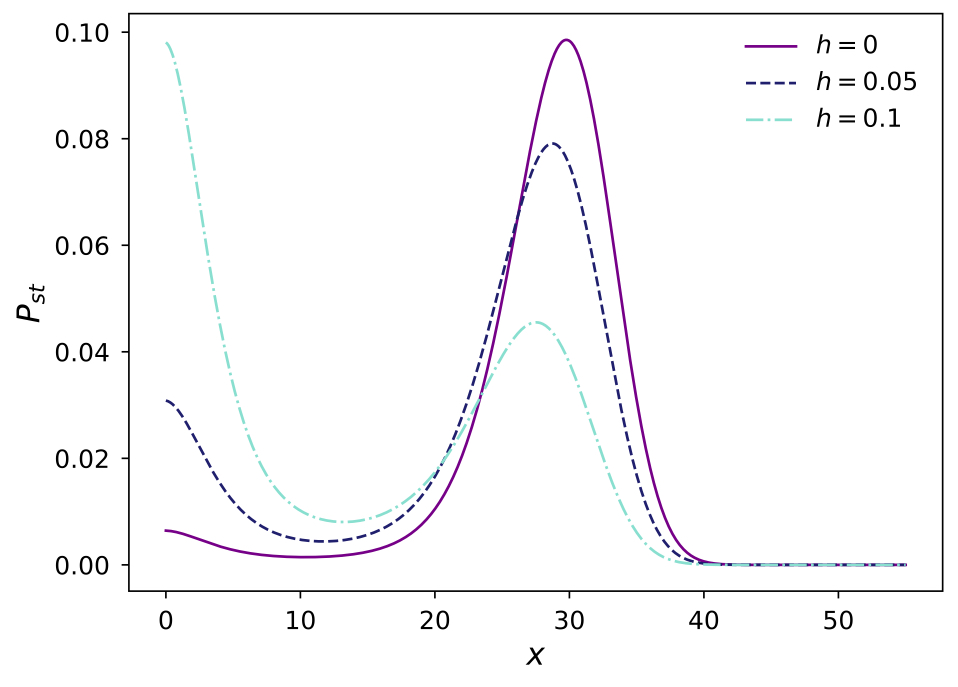}%
\caption{$P_{st}$ as a function of $x$ for different values of $h$ when $r=0.01$, $K=30$, $m=10$, $b_1=5$, $b_2=0.1$, $D=0.001$, $\lambda=0.5$ and $\alpha= 0.9$.}
\end{figure}

\begin{figure}[ht]
\includegraphics[scale=0.24]{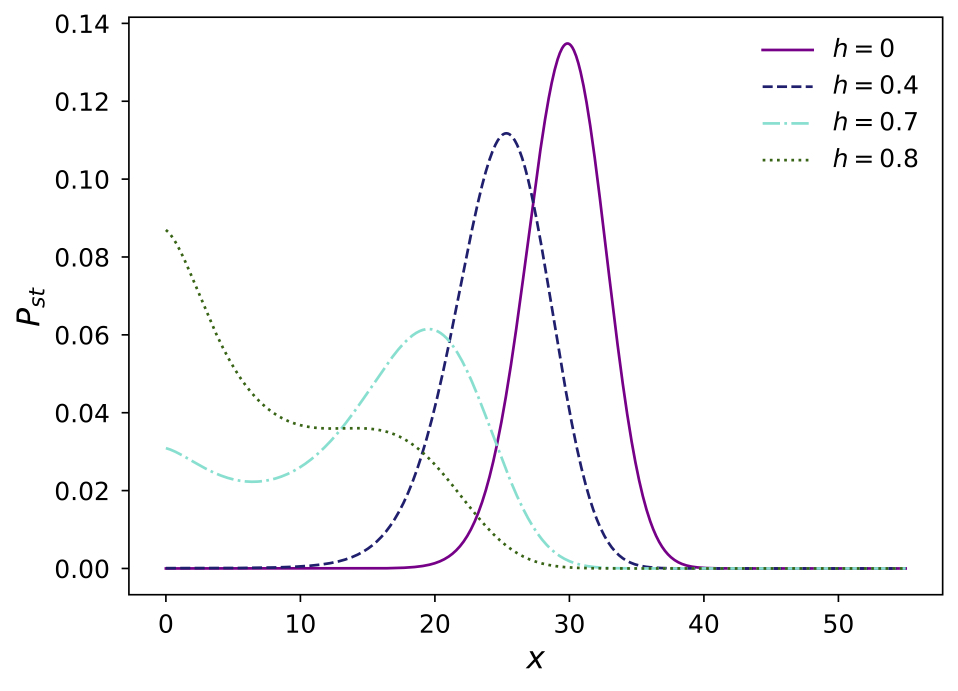}%
\caption{$P_{st}$ as a function of $x$ for different values of $h$ when $r=0.01$, $K= 30$, $m=-10$, $b_1=5$,  $b_2=0.1$, $D=0.001$, $\lambda=0.9$ and $\alpha= 0.9$.}
\end{figure}

\newpage

\section{Stationary Probability Distribution of Population Subject to Colored Noise and Harvesting} \label{sec4}

In this section, we consider cubic population model containing Allee effect and subject to colored noise. We write Langevin equation as follows:

\begin{eqnarray}
\frac{{dx}}{{dt}} =&&  - \frac{r}{K}{x^3} + r\left( {1 + \frac{m}{K}} \right){x^2} \nonumber\\
&&- rmx -H(x) + x\xi (t) + \eta (t)
\end{eqnarray}

where $\xi(t)$ and $\eta(t)$ are correlated Gaussian colored noises with zero mean. Their statistical properties are given as:

\begin{subequations}
\begin{align}
&\left\langle {\xi (t)} \right\rangle  = \left\langle {\eta (t)} \right\rangle  = 0\\
&\left\langle {\xi (t)\xi (t')} \right\rangle =
\frac{Q}{\tau_1}\exp(-\frac{|t-t'|}{\tau_1})\\
&\left\langle {\eta (t)\eta (t')} \right\rangle  = \frac{P}{\tau_2}\exp(-\frac{|t-t'|}{\tau_2})\\
&\left\langle {\xi (t)\eta (t')} \right\rangle = \frac{\lambda'\sqrt{PQ}}{\tau_3}\exp(-\frac{|t-t'|}{\tau_3})\\
&\left\langle {\xi (t)\eta (t')} \right\rangle  = \left\langle {\eta (t)\xi (t')} \right\rangle
\end{align}
\end{subequations}
Here, $Q$ and $P$ are the strength of Gaussian colored noises $\xi(t)$ and $\eta(t)$. $\lambda'$ is the degree of correlation between noises. $\tau_1$ with $\tau_2$ are the self correlation times of $\xi(t)$ and $\eta(t)$ with $\tau_3$ being the cross correlation time between the noises. \par
Time evolution of probability distribution, $P(x,t)$ which is the probability of having $x$ individuals at time $t$ is given by the Approximate Fokker-Planck equation (AFPE). We write AFPE by using the procedure given in \cite{Liang2004, Da-Jin1994,Sancho1982,Novikov1965} as follows:

\begin{eqnarray}
\frac{\partial }{{\partial t}}P(x,t) =&&  - \frac{\partial }{{\partial x}}\left[ {A(x)P(x,t)} \right] \nonumber \\
&&+ \frac{{{\partial ^2}}}{{\partial {x^2}}}\left[ {B(x)P(x,t)} \right]
\end{eqnarray}

\begin{subequations}
\begin{eqnarray}
A(x) = && f(x)+\frac{Qx}{1-\tau_1 f'(x_s)} \nonumber \\
&&+\frac{\lambda\sqrt{PQ}}{1-\tau_3f'(x_s)}
\end{eqnarray}
\begin{eqnarray}
B(x) = && \frac{Qx^2}{1-\tau_1 f'(x_s)} +2\frac{\lambda\sqrt{PQ}x}{1-\tau_3 f'(x_s)} \nonumber \\
&&+ \frac{P}{1-\tau_2f'(x_s)}
\end{eqnarray}
\begin{eqnarray}
f(x) = && -\frac{r}{K}x^3 +r\Big(1+\frac{m}{K}\Big)x^2 \nonumber \\
&&-rmx -H(x) 
\end{eqnarray}
\end{subequations}

Here, $x_s$ is the stable point of $f(x)$ \cite{Liang2004,Fronzoni1986}. AFPE is valid for $1-\tau_i f'(x_s)>0$ for $i=1,2,3$ and $\tau_1\neq0$, $\tau_2\neq0$, $\tau_3\neq0$.

Stationary solution of the given Approximate Fokker-Planck equation obtained as follows:
\begin{eqnarray}
{P_{st}} = \frac{N}{{B(x)}}\exp \left[ {\int_{}^x {\frac{{A(x')}}{{B(x')}}dx'} } \right]
\end{eqnarray}
Where $N$ is the normalization constant.\par

\subsection{Linear Harvesting}
Without considering the correlation between noises and harvesting, model 4.1 with 4.2 investigated and effects of $Q, P, \tau_1$ and $\tau_2$ on SPD discussed in Ref. \cite{Yang2019a}. Thus, we first look at the effects the of degree of correlation between noises and cross correlation time without considering the harvesting. In Fig. 26, change of SPD for different values of $\lambda'$ given for strong Allee effect case. When $\lambda'=0$, there is no correlation between noises and probability of population to be around carrying capacity is the highest. For the increasing values of $\lambda'$, probability of population to be around extinction point, $0$, is also increased while probability to be around carrying capacity is decreased. Fig. 27 shows effect of changing values of $\lambda'$ for weak Allee effect case. In this situation, shape of SPD is unimodal on the contrary of previous strong Allee effect case. However, probability of population to be around carrying capacity decreased for the increasing values of $\lambda'$ again. For the increasing values of $\lambda'$ we do not see a significant change of SPD around extinction point.\par
\begin{figure}[ht]
\includegraphics[scale=0.24]{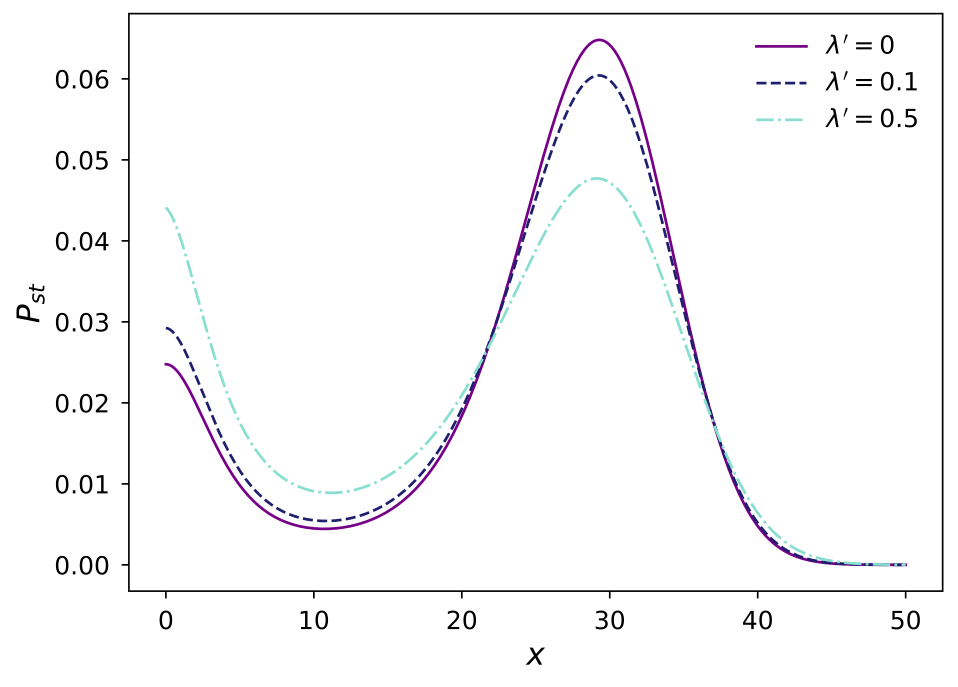}%
\caption{$P_{st}$ as a function of $x$ for different values of $\lambda'$ when $r=0.01$, $K= 30$, $m=10$, $Q=0.005$, $P=0.9$, $\tau_1=0.5$, $\tau_2=0.5$ and $\tau_3=0.5$.}
\end{figure}

\begin{figure}[ht]
\includegraphics[scale=0.24]{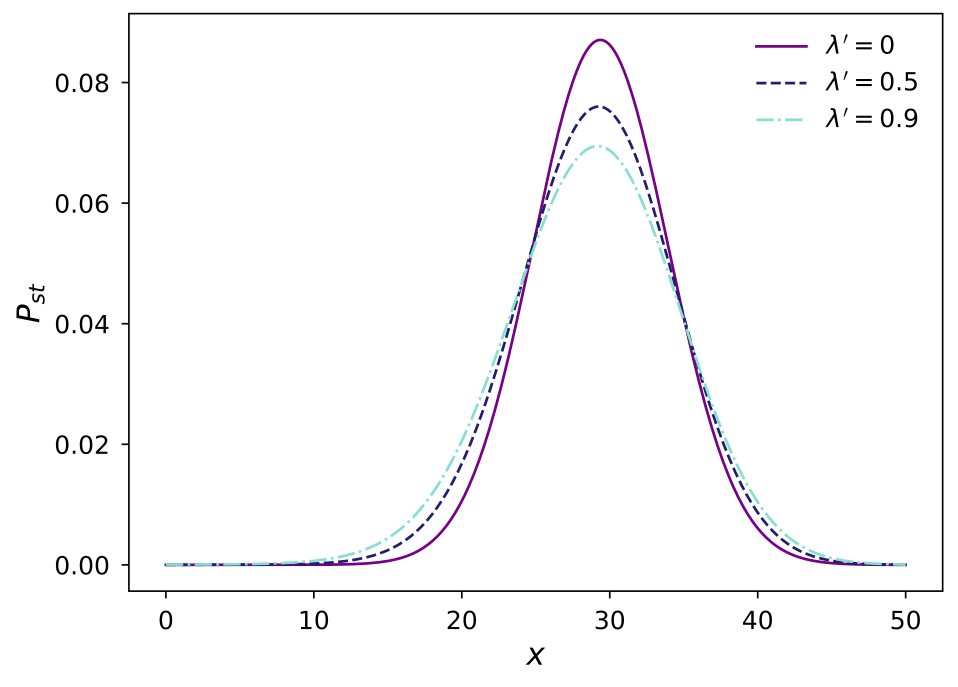}%
\caption{$P_{st}$ as a function of $x$ for different values of $\lambda'$ when $r=0.01$, $K= 30$, $m=-10$, $Q=0.01$, $P=0.9$, $\tau_1=0.5$, $\tau_2=0.5$ and $\tau_3=0.5$.}
\end{figure}

Fig. 28 shows change of SPD for different values of cross-correlation time, $\tau_3$, for strong Allee effect case. With increasing values of $\tau_3$, bimodal shape of SPD is preserved and probability of population to extinct is reduced. Also, as the cross-correlation time increased, probability of population to be around carrying capacity increased. In Fig. 29 it is given for weak Allee effect case. Increase of cross-correlation time does not change unimodal structure of SPD but the probability of population to be around carrying capacity is increased whereas probability to extinct does not show any significant change.

\begin{figure}[ht]
\includegraphics[scale=0.24]{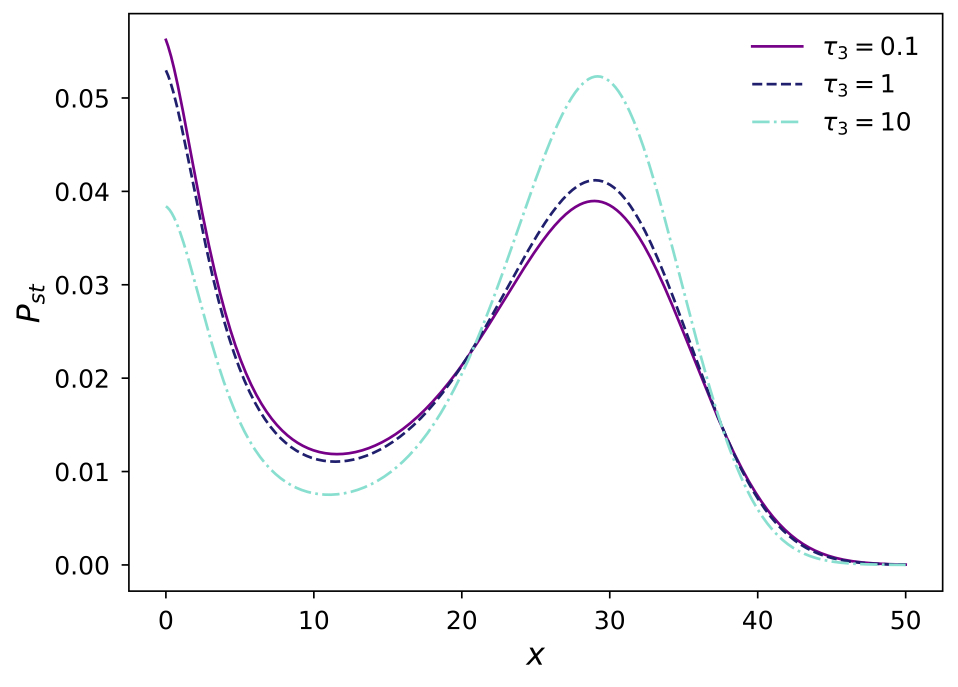}%
\caption{$P_{st}$ as a function of $x$ for different values of $\tau_3$ when $r=0.01$, $K= 30$, $m=10$, $Q=0.005$, $P=0.9$, $\lambda'=0.9$, $\tau_1=0.5$ and $\tau_2=0.5$.}
\end{figure}

\begin{figure}[ht]
\includegraphics[scale=0.24]{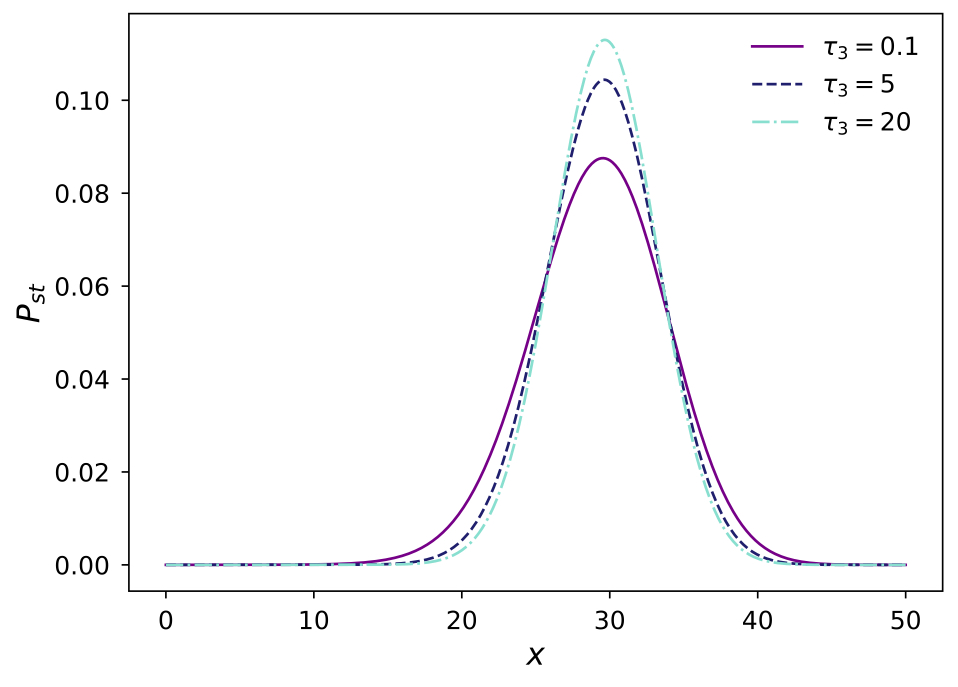}%
\caption{$P_{st}$ as a function of $x$ for different values of $\tau_3$ when $r=0.01$, $K= 30$, $m=-10$, $Q=0.005$, $P=0.9$, $\lambda'=0.9$, $\tau_1=0.5$ and $\tau_2=0.5$}
\end{figure}

\begin{figure}[ht]
\includegraphics[scale=0.24]{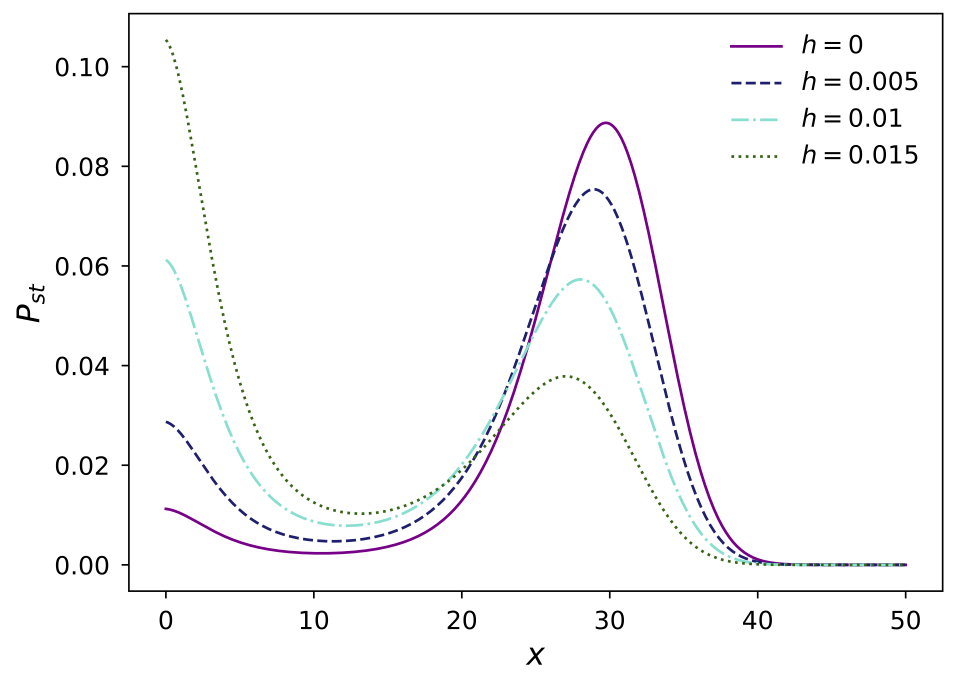}%
\caption{$P_{st}$ as a function of $x$ for different values of $h$ when $r=0.01$, $K= 30$, $m=10$, $Q=0.001$, $P=0.9$, $\lambda'=0.9$, $\tau_1=0.5$, $\tau_2=0.5$ and $\tau_3=0.5$.}
\end{figure}

\begin{figure}[ht]
\includegraphics[scale=0.24]{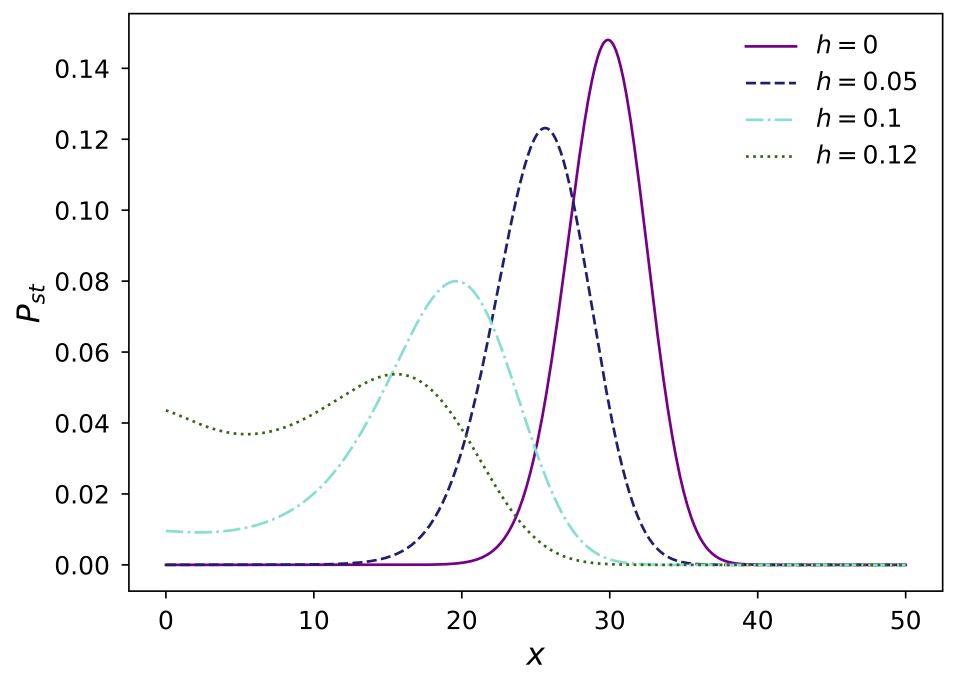}%
\caption{$P_{st}$ as a function of $x$ for different values of $h$ when $r=0.01$, $K= 30$, $m=-10$, $Q=0.001$, $P=0.9$, $\lambda'=0.9$, $\tau_1=0.5$, $\tau_2=0.5$ and $\tau_3=0.5$.}
\end{figure}

We consider harvesting function as a linear function of $x$ in the form of, $H(x)=hx$. In Fig. 30, change of SPD with respect to different values of $h$ for strong Allee effect case given. As $h$ is increased, probability of population to extinct is increased while the unimodal shape of SPD is preserved. Because presence of linear harvesting function cause stationary points to change, carrying capacity and Allee threshold get close to each other as $h$ is increasing. Change of SPD for different values of $h$ in the case of weak Allee effect is given in Fig. 31. As $h$ increases, unimodal shape of SPD turns to bimodal, and probability of population to extinct increases. As the shape of the SPD turns to bimodal, weak Allee effect turns to strong Allee effect because of the effect of increasing harvesting function on stationary points of the system.

\subsection{Nonlinear Harvesting (Holling type-II)}
Here, we consider harvesting function as a nonlinear function of $x$ in the form of $H(x)=hx/(b_1+b_2x)$. In Fig. 32, for strong Allee effect case, SPD change with respect to different values of $h$ given. As the values of $h$ increase, carrying capacity and Allee threshold changes so that they get close to each other while the height of the peak situated on the carrying capacity gets smaller and probability of population to be around extinction point gets bigger. Change of SPD with respect to $h$ given in Fig. 33 is for the weak Allee effect case. We see that as $h$ increases, unimodal shape of SPD turns to bimodal and weak Allee effect turns to strong Allee effect. Also, probability of population extinction is increased with increasing values of $h$.

\begin{figure}[ht]
\includegraphics[scale=0.24]{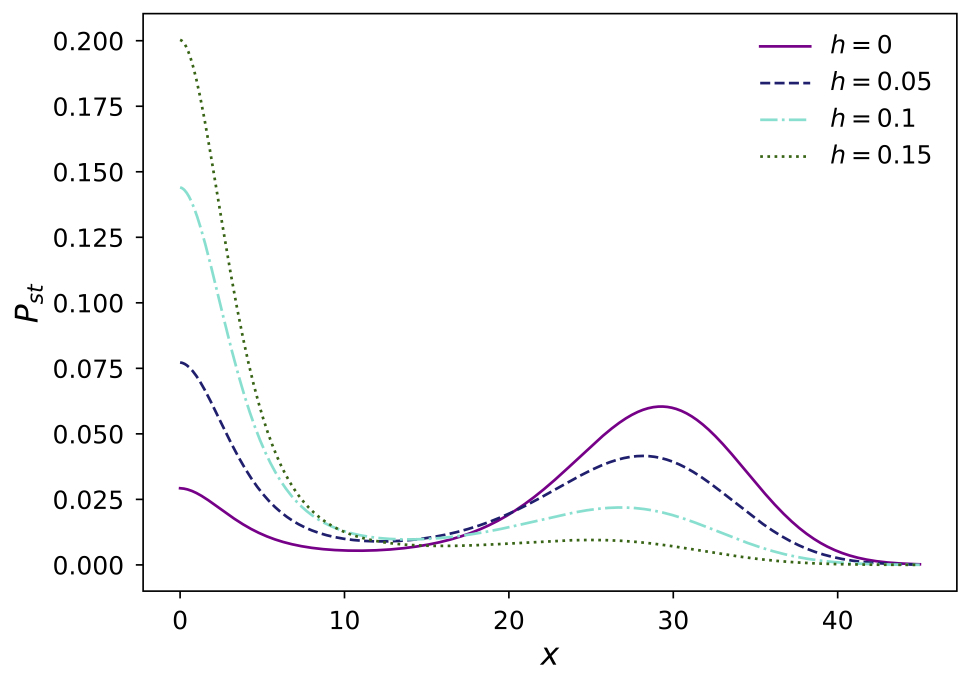}%
\caption{$P_{st}$ as a function of $x$ for different values of $h$ when $r=0.01$, $K= 30$, $m=10$, $b_1=5$, $b_2=0.1$ $Q=0.005$, $P=0.9$, $\lambda'=0.1$, $\tau_1=0.5$, $\tau_2=0.5$ and $\tau_3=0.5$.}
\end{figure}

\begin{figure}[ht]
\includegraphics[scale=0.24]{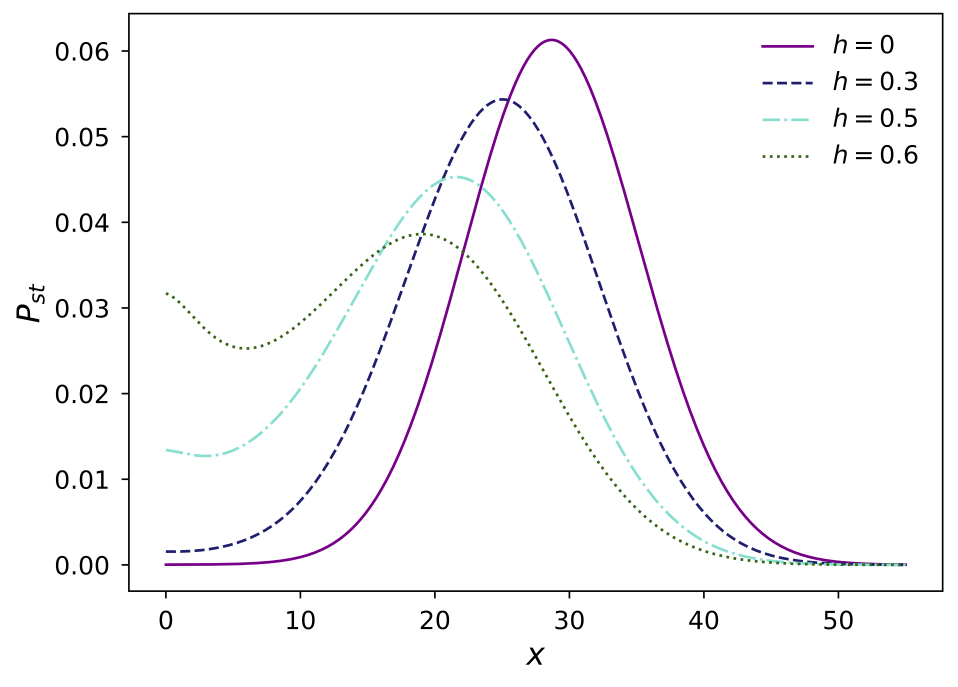}%
\caption{$P_{st}$ as a function of $x$ for different values of $h$ when $r=0.01$, $K= 30$, $m=-10$, $b_1=5$, $b_2=0.1$ $Q=0.02$, $P=0.9$, $\lambda'=0.1$, $\tau_1=0.5$, $\tau_2=0.5$ and $\tau_3=0.5$.}
\end{figure}
\newpage
\section{Conclusion} \label{sec5}
We discussed the effect of harvesting on a population model under Allee effect. We made a detailed examination of the effect of harvesting by writing a cubic deterministic model under weak or strong Allee effect depending on the size of the parameters. When the harvesting is represented by a constant in the form $H(x)=H$, a population under weak Allee effect immediately starts to show strong Allee effect. As the harvesting term increases, the value of the Allee threshold increases, while the carrying capacity of the population decreases. In the deterministic model we made a similar analysis for the case where harvesting is represented by a function in the form of $H(x)=H_1x^2 +H_2x$. We discussed weak and strong Allee effect transitions, carrying capacity and Allee threshold change according to the change of harvesting function.\par

We developed a stochastic description of the cubic model. We have written Langevin equations of two population models subject to Gaussian white and Gaussian colored noises, respectively, and obtained Fokker-Planck equation for the population subject to Gaussian white noise and Approximate Fokker-Planck equation for the population subject to colored noise. Thus, we calculated the stationary probability distributions of populations in the presence of harvesting for populations subject to either Gaussian white or Gaussain colored noise. For the population model subject to Gaussian white noise, we discussed the effect of multiplicative noise strength, $D$, additive noise strength, $\alpha$, and degree of correlation between noises, $\lambda$, on stationary probability distribution. In the case of population model subject to Gaussian colored noise, we discussed the effect of degree of correlation between noises, $\lambda’$ and cross correlation time, $\tau_3$, on stationary probability distribution.  We have considered two types of harvesting functions, linear and Holling type-II, and discussed how their change effect stationary probability distributions of populations subject to either Gaussian white or Gaussian colored noise. \par
We conclude that, intensity of harvesting and the strength of noise being small, increase the probability of population to be around carrying capacity whereas increase of the intensity of harvesting and the strength of noise reduce probability of population size to be around carrying capacity and increase the probability of extinction. Increase of the cross-correlation time of colored noises, $\tau_3$, increase probability of population survival. On the contrary, degree of correlation between colored noises, $\lambda’$, and degree of correlation between white noises, $\lambda$, increase the probability of extinction. Even if the white and colored noise strengths and degree of correlations have similar effect on stationary probability distributions, they change stationary probability distributions completely differently. \par
In fact, there are some issues about examining the model stochastically by adding a white noise to the model. First of all, white noise with zero autocorrelation should not be seen as a realistic noise. Rather than seeing the white noise we add to the model as a real noise, we consider it as a theoretical object that can be interpreted over time. Therefore, colored noise with nonzero autocorrelation offers us a more realistic stochastic model than the white noise. However, we need to highlight some of the limitations of our study. For example, environmental noise is not effective on the population, but on processes on individuals. Although in this study we have followed the tradition of obtaining a stochastic model by adding noise to the population model, we are aware that the stochastic model we put forward with this approach has only phenomenological consequences. However, for a more effective analysis of our results, it may be the subject of our future studies to enrich the analysis by developing a model that takes into account the effect of environmental noise on individual level processes in a population model showing Allee effect.

\section{Acknowledgements}
We thank Dr. Özgür Gürcan for providing suggestions for the improvement of the text.

\section{Conflict of Interest}
The authors declare that they have no conflicts of interest.

\newpage

\bibliography{main.bib}
\bibliographystyle{unsrt}

\end{document}

%% file: preamble.tex
\usepackage{amsthm}
\usepackage{mathtools}
\usepackage{physics}
\usepackage{xcolor}
\usepackage{graphicx}
\usepackage[left=23mm,right=23mm,top=35mm,columnsep=15pt]{geometry} 
\usepackage{adjustbox}
\usepackage{placeins}
\usepackage[T1]{fontenc}
\usepackage{lipsum}
\usepackage{csquotes}